\documentclass[a4paper, amsfonts, amssymb, amsmath, showkeys, floatfix, nofootinbib, twoside, superscriptaddress, aps, prl, reprint]{revtex4-2}
\usepackage[english]{babel}
\usepackage[utf8]{inputenc}
\usepackage{mathtools}
\usepackage{hyperref}

\begin{document}

\title{Universal Defect Statistics in Reverse Quenches}
\thanks{The work presented here is a part of \href{https://10.11588/heidok.00038091}{E. J. Braun, PhD thesis Heidelberg University, 2026}}

\author{Eduard J. Braun}
\email{ebraun@physi.uni-heidelberg.de}
\affiliation{Physikalisches Institut, Universit\"at Heidelberg, Im Neuenheimer Feld 226, 69120 Heidelberg, Germany}
\author{Daniel Rubin}
\affiliation{Physikalisches Institut, Universit\"at Heidelberg, Im Neuenheimer Feld 226, 69120 Heidelberg, Germany}
\author{Margaux Cartier}
\affiliation{Physikalisches Institut, Universit\"at Heidelberg, Im Neuenheimer Feld 226, 69120 Heidelberg, Germany}
\affiliation{Laboratoire de Physique de l’Ecole normale supérieure, ENS, Université PSL, CNRS,
Sorbonne Université, Université Paris Cité, F-75005 Paris, France}
\author{Gerhard~Z\"urn}
\affiliation{Physikalisches Institut, Universit\"at Heidelberg, Im Neuenheimer Feld 226, 69120 Heidelberg, Germany}
\author{Matthias~Weidem\"uller}
\email{weidemueller@uni-heidelberg.de}
\affiliation{Physikalisches Institut, Universit\"at Heidelberg, Im Neuenheimer Feld 226, 69120 Heidelberg, Germany}

\date{\today}

\begin{abstract}

We extend the generalized Kibble–Zurek mechanism (gKZM) to reverse quench protocols using a stochastic description of defect formation finding that the mean defect density after a reverse quench equals twice the forward quench defect-number variance per unit volume. This relation, universally valid within the stochastic framework of the gKZM, makes the forward quench variance accessible through a reverse quench defect density measurement. We verify the equality analytically and numerically in the uniform one-dimensional transverse-field Ising model, and numerically in its bond-disordered variant, including a regime where defect scaling departs from a pure power-law in quench time. Our result also suggests an experimental route to probing quantum critical behavior: reverse quenches that begin and end in the paramagnetic phase may allow to reveal critical exponents using global magnetic-field control and magnetization measurements, without resolving individual defects.

\end{abstract}

\maketitle

The Kibble-Zurek mechanism (KZM) relates the formation of defects during a passage through a continuous phase transition to the equilibrium critical exponents of that transition \cite{kibble_topology_1976,zurek_cosmological_1985,zurek_cosmological_1996}. The generalized KZM (gKZM) extends this picture beyond the mean defect density by describing the full counting statistics of defects, whose cumulants can provide access to critical scaling \cite{del_campo_universal_2018,gomez-ruiz_full_2020,bando_probing_2020,cui_experimentally_2020,mayo_distribution_2021,zhang_work_2022,grabarits_driving_2025,wang_quantum_2026}.

Most KZM studies consider a forward quench: a single passage through the critical point. Reverse quenches, which cross the critical point twice, have also been studied in one-dimensional systems, including the XY, Kitaev, and transverse-field Ising chains \cite{mukherjee_defect_2008,divakaran_reverse_2009,quan_testing_2010,kou_interferometry_2022}. These works establish that reverse quenches can produce characteristic defect scaling, but the results have largely been obtained in specific models. A general stochastic account of reverse-quench defect statistics, and in particular of their relation to forward-quench fluctuations, remains to be established.

In this Letter, we extend the stochastic generalized Kibble--Zurek mechanism to reverse quenches and show that the mean defect density after a reverse quench equals twice the defect-number variance per unit volume after the corresponding forward quench. This relation is universal within the stochastic gKZM framework. We verify it in uniform and bond-disordered one-dimensional TFIMs, the latter providing a qualitatively distinct test in a system governed by an infinite-randomness fixed point.

\begin{figure}[t]
    \centering
    \includegraphics[width=0.9\linewidth]{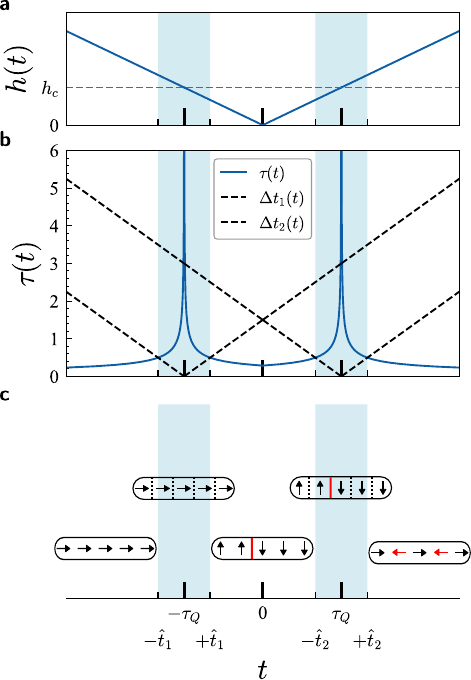}
    \caption{ (a) Change of the magnetic field h (blue solid line) as a function of time \(t\). Critical field \(h_c\) is indicated by a red dotted line, reached at times \(\pm \tau_Q\). 
    (b) Comparison of relevant time scales. Instantaneous relaxation time \(\tau(t)\) (blue solid) and time difference to the time when the critical point is reached \(\Delta t_1(t)\) and \(\Delta t_2(t)\) (black dotted). The frozen region according to the adiabatic impulse approximation (see details in main text) is shaded in cyan throughout all panels.
    (c) Expected spin configurations (color arrows) for a 1D Ising ferromagnet. Protodomain walls are indicated by black dashed lines in the cyan shaded regions. Domain walls are shown by red solid lines. In the paramagnetic phase, these are adiabatically transferred into spin flips (red arrows).
    }
    \label{fig:KZM}
\end{figure}

\textit{Framework and system.--} We consider a system undergoing a continuous phase transition controlled by a parameter \(h\). 
Choosing units such that \(h_c=1\), the correlation length \(\xi\) and characteristic relaxation time \(\tau\) scale near criticality as
\begin{align}
    \xi(h) &= \xi_0 |h-1|^{-\nu}, \\
    \tau(h) &= \tau_0 |h-1|^{-z\nu},
\end{align}
where \(\nu\) and \(z\) are the correlation-length and dynamical critical exponents, respectively, and \(\xi_0\) and \(\tau_0\) are nonuniversal scaling amplitudes. As is the case in most (quantum) phase transitions, we assume \(z\) and \(\nu\) to be equal on both sides of the critical point \cite{sachdev_quantum_2011,del_campo_universality_2014}.

We study the reverse quench protocol shown in Fig. \ref{fig:KZM}a,
\begin{equation}
    h(t)= \frac{|t|}{\tau_Q},
    \qquad -h_0 \tau_Q \le t \le h_0 \tau_Q,
    \label{eq:reverse_protocol}
\end{equation}
with \(h_0\gg1\). The system starts and ends in the \(h>1\) phase, crossing the critical point \(h=1\) twice, namely at \(t=\pm\tau_Q\), and passing through the \(h<1\) phase. The quench time \(\tau_Q\) sets the ramp rate which is the same at each crossing.

\textit{General derivation.--} The first half of the protocol, \(t<0\), crosses the critical point once and is therefore a forward quench in the sense used in Ref.~\cite{del_campo_universal_2018}, where the gKZM was introduced. Here, a forward quench refers to a single passage through the critical point and may proceed in either direction. The system begins in thermal equilibrium (or, at zero temperature, in its ground state) at \(h=h_0\).

As shown in Fig. \ref{fig:KZM}b, the forward quench defect density \(n_{\mathrm{def}}\) and its variance follow from the competition between the relaxation time and the time to the first critical crossing, \(t_c^{(1)}=-\tau_Q\). Defining \(\Delta t_1(t)=|t-t_c^{(1)}|\), the freeze-out times \(t_c^{(1)}\pm\hat{t}_1\) are set by \(\tau\left(t_c^{(1)}\pm\hat{t}_1\right)=\Delta t_1\left(t_c^{(1)}\pm\hat{t}_1\right)\). Between these times, the system cannot follow the drive adiabatically; in the adiabatic-impulse approximation,  which has been experimentally verified on several platforms \cite{xu_quantum_2014,navon_critical_2015,cui_experimental_2016,gong_simulating_2016}, its correlation length remains fixed at \(\hat{\xi}_1\equiv\xi[h(t_c^{(1)}-\hat{t}_1)]\).
 
The gKZM \cite{del_campo_universal_2018} describes defect formation stochastically. During freeze-out, the correlation length is fixed at \(\hat{\xi} = \xi[t_c^{(1)} - \hat{t}_1(\tau_Q)]\), defining equally sized protodomains as a mathematical partition. Each protodomain wall is assumed to become a real domain wall with an independent probability \(p\); as the system evolves adiabatically in the \(h < 1\) phase, these domain walls are transferred to defects. We therefore use the terms domain wall and defect interchangeably. 
For point defects in a system in \(D\) dimensions of volume \(V\), the number of possible formation sites is \(\mathcal{N} = V/(f\hat{\xi}^{D})\), where \(f\) is a geometric factor \cite{gomez-ruiz_full_2020}. The defect count follows a binomial distribution, and the expectation value of the defect density reads
\begin{equation}
    n_{\mathrm{def}}(\tau_Q)
    = \frac{\mathcal{N} p}{V}
    = \frac{p}{f\hat{\xi}^{D}}
    = \frac{p}{f\xi_0^{D}}
      \left(\frac{\tau_Q}{\tau_0}\right)^{-\mu},
    \quad
    \mu=\frac{D\nu}{1+z\nu},
    \label{eq:gKZM}
\end{equation}
consistent with the KZM scaling derived using only the adiabatic-impulse approximation \cite{del_campo_universality_2014}.

The gKZM also predicts the higher-order cumulants of the defect number from the binomial distribution. Here, we define the defect-number variance per unit volume as \(\mathrm{Var}(N_{\mathrm{def}})/V\), where \(N_{\mathrm{def}}\) is the number of defects. It is given by
\begin{equation}
\begin{split}
    \mathrm{Var}(n_{\mathrm{def}})
    &= \frac{p(1-p)}{f\hat{\xi}^D}
    = \frac{p(1-p)}{f\xi_0^D}
      \left(\frac{\tau_Q}{\tau_0}\right)^{-\mu},\\
      &= (1-p) n_\mathrm{def}(\tau_Q)
       \label{eq:gKZM_var}
\end{split}
\end{equation}
Thus, the variance scales with the same exponent \(\mu\) as \(n_\mathrm{def}\), and their ratio is time-independent and only determined by the defect formation probability \(p\).

For the second half of the protocol (\(t > 0\)), the system crosses the critical point again at \(t_c^{(2)} = \tau_Q\). After following the drive adiabatically in the \(h < 1\) phase at \(t = 0\), it undergoes a second forward quench—that is, a single passage through the critical point, now in the opposite direction. Since the critical exponents and ramp rates are the same at both crossings, the freeze-out times \(t_c^{(2)} \pm \hat{t}_2\), determined by \(\tau\!\left(t_c^{(2)} \pm \hat{t}_2\right) = \Delta t_2\!\left(t_c^{(2)} \pm \hat{t}_2\right)\), with \(\Delta t_2(t) = |t - t_c^{(2)}|\), yield the same freeze-out length as at the first crossing. We therefore use the same effective protodomain partition for both passages.

In the stochastic gKZM, each effective protodomain site forms a defect during a single passage with probability \(p\), independently of the other sites \cite{del_campo_universal_2018}. Since the two passages have the same critical exponents and ramp rate, we take the defect formation probability at the second passage to be the same \(p\) consistent with \cite{dziarmaga_dynamics_2005}. If no defect formed during the first passage, the second passage therefore produces one with probability \(p\), as in an ordinary forward quench. The additional assumption introduced in this Letter concerns sites where a defect did form during the first passage: we assume that the second passage destroys that defect with the same probability \(p\). Assuming independent events at the two passages, a defect survives the full protocol either by forming at the first passage and surviving the second, or by forming only at the second. Each pathway has probability \(p(1-p)\), so the total survival probability is \(2p(1-p)\). This resulting probability is consistent with the model-specific calculation of Ref. \cite{yan_nonadiabatic_2021} for an Ising chain with two spins per unit cell driven across two critical points with identical critical exponents, a setting that can be mapped to the reverse-quench protocol considered here.

Treating defect formation at the effective protodomain sites as independent, as in the gKZM, the reverse quench defect density reads
\begin{equation}
    \begin{split}
        n_\mathrm{def}^\mathrm{rev} &= \frac{\mathcal{N}}{V} 2p\left(1-p \right) = 2\frac{p\left(1-p\right)}{f \xi_0^D} \left(\frac{\tau_Q}{\tau_0} \right)^{-\mu},\\
        &= 2\mathrm{Var}\left(n_\mathrm{def}\right).
    \end{split}
\label{eq:rev_scaling}
\end{equation}
This result connects an expectation value in one phase to a fluctuation in another through a dynamical protocol. The stochastic picture also explains why the reverse quench defect density scales with the same critical exponent \(\mu\), consistent with model-specific calculations for several one-dimensional systems \cite{mukherjee_defect_2008,divakaran_reverse_2009,quan_testing_2010,kou_interferometry_2022}.

\textit{Analytic benchmark.--} We benchmark our findings analytically studying the dimensionless TFIM Hamiltonian 
\begin{equation}
    H = -\sum_{i=1}^N J_i \sigma_i^z \sigma_{i+1}^z -h \sum_{i=1}^N \sigma_i^x,
    \label{eq:H_Ising}
\end{equation}
where \(N\) is the number of spins, \(\sigma_i^\alpha\) a Pauli-\(\alpha\) operator acting on spin \(i\), \(J_i>0\) the coupling constant between spins \(i\) and \(i+1\), and \(h\) an external transverse magnetic field. Throughout this article, we apply periodic boundary conditions, i.e. \(\sigma^\alpha_{N+1}=\sigma_{1}^\alpha\). In the case of the uniform transverse field Ising model (TFIM) which we consider now, \(J_i = 1\) for all \(i\) and a quantum phase transition at \(h_c=1\) occurs with critical exponents \(z=\nu=1\) and \(\mu=1/2\) \cite{sachdev_quantum_2011}. For convenience, we restrict without loss of generality to the case of an even number of spins \cite{he_boundary_2017}.

As detailed in Ref. \cite{dziarmaga_dynamics_2005}, after a Jordan-Wigner and a Fourier transformation, this Hamiltonian can be brought into a sum of independent two-level systems undergoing a Landau-Zener passage with gap \(\Delta_k = \left(4\tau_Q\sin^2(k)\right)^{-1}\). For slow quenches (\(\tau_Q\gg1\)), and setting \(h_0 \rightarrow \infty\), only long-wavelength modes can be excited, and the excitation probability for mode \(k\) is given as
\begin{equation}
    p_k = e^{-\frac{\pi}{2\Delta_k}}.
    \label{eq:transition_prob}
\end{equation}
As a consequence, the forward quench defect density reads in the thermodynamic limit \(N\rightarrow\infty\)
\begin{equation}
\begin{split}
   n_\mathrm{def} = \frac{1}{2\pi}\int_{-\pi}^\pi p_k dk\approx \frac{1}{2\pi}\int_{-\infty}^\infty p_k dk = \frac{1}{2\pi} \frac{1}{\sqrt{2\tau_Q}}\sim \tau_Q^{-\frac{1}{2}}.
   \end{split}
    \label{eq:analytic_ndef}
\end{equation}

Using the structure of the Landau-Zener crossings, as detailed in the Supplemental Material \cite{supplement}, in the same limit the defect density variance can be calculated as 
\begin{equation}
    \mathrm{Var}(n_\mathrm{def}) = \frac{1}{2\pi} \int_{-\pi}^\pi \left(p_k - p_k^2\right) dk= \left(1-\frac{1}{\sqrt{2}}\right) n_\mathrm{def},
\label{eq:analytic_var_ndef}
\end{equation}
yielding the same result as previous calculations \cite{del_campo_universal_2018,cui_experimentally_2020}. A comparison to Eq. \ref{eq:gKZM_var} yields a defect formation probability \(p=1/\sqrt{2}\).

\begin{figure*}[t]
    \centering
    \includegraphics[width=\linewidth]{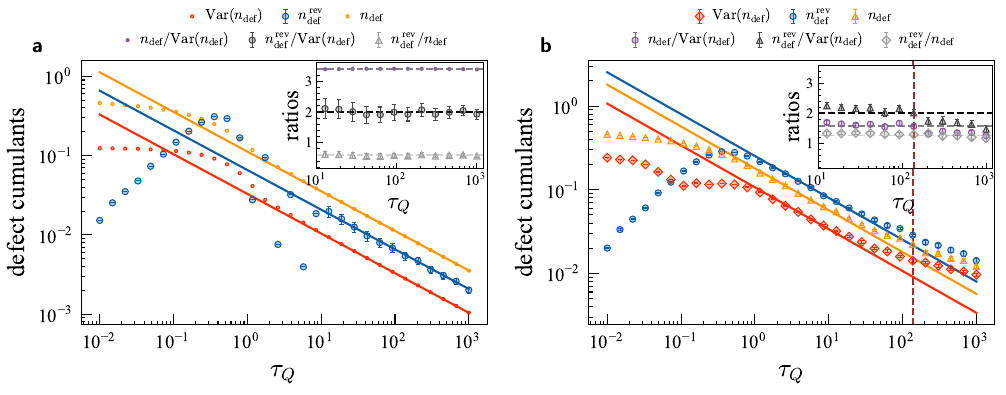}
    
    \caption{Defect density after forward (orange) and reverse (blue) quenches, and defect density variance after a forward quench (red), versus quench time \(\tau_Q\) for the uniform (a) and bond-disordered (b) TFIMs. Solid lines in (a) show the power-law predictions from Eqs. \ref{eq:analytic_ndef}, \ref{eq:analytic_revdef}, and \ref{eq:analytic_var_ndef}. In (b), the lines illustrate the power-law dependence \(n_\mathrm{def}=C/(2\pi\sqrt{2\tau_Q})\), with \(C=1.6\), \(0.95\), and \(2.25\) for the forward quench defect density, forward quench defect density variance, and reverse quench defect density, respectively. Error bars show the standard error of the mean, calculated from 11 evenly spaced points spanning approximately two Stückelberg oscillations in the uniform model and from 20 disorder realizations in the bond-disordered model (see Supplemental Material \cite{supplement}). Insets show, in violet, gray, and light gray, respectively, the ratios of the forward quench and reverse quench defect densities to the forward quench variance, and the ratio of the forward to the reverse quench defect density. The black, gray, and light-gray dashed lines mark \(2\), the value \(1/(1-p)\), and \(2-2p\), respectively, where \(p = 1/\sqrt{2}\) for the uniform TFIM and \(p = 0.37\) for the bond-disordered TFIM.}
    
    \label{fig:defects}
\end{figure*}

An analytical solution for the reverse quench defect density at arbitrary \(\tau_Q\) was derived in Ref. \cite{kou_interferometry_2022}. Here, we use a simpler approach to determine its behavior at large \(\tau_Q\) analytically, describing the double passage in terms of Stückelberg oscillations. In this regime, the excitation probability is given by
\begin{equation}
    p_k (\tau_Q) = 4 e^{-\frac{\pi}{2\Delta_k}}\left(1-e^{-\frac{\pi}{2\Delta_k}}\right)\sin^2(\Phi_\mathrm{St}),
    \label{eq:Stueckelberg}
\end{equation}
where we calculate the Stückelberg phase \cite{stueckelberg_theorie_1932,shevchenko_landauzenerstuckelberg_2010,ivakhnenko_nonadiabatic_2023} \(\Phi_\mathrm{St}\approx2\tau_Q\), as shown in more detail in the Supplemental Material \cite{supplement}. Thus, in the thermodynamic limit averaging over one period of these oscillation yields

\begin{equation}
\begin{split}
    n_\mathrm{def}^\mathrm{rev} &= \frac{1}{2\pi} \int_{-\pi}^\pi \overline{p_k} dk
    \approx \frac{1}{2\pi} \int_{-\pi}^\pi 2 e^{-\frac{\pi}{2\Delta_k}}\left(1-e^{-\frac{\pi}{2\Delta_k}}\right) dk,\\
    &= \left(2-\sqrt{2}\right) n_\mathrm{def} = 2\mathrm{Var}\left(n_\mathrm{def}\right),
\end{split}
\label{eq:analytic_revdef}
\end{equation}
where the overbar denotes an average over one oscillation period. We derive that in the limit of large \(\tau_Q\), the reverse quench defect density equals twice the forward quench variance.

\textit{Numerical benchmarks.—} We benchmark Eqs. \ref{eq:gKZM} and \ref{eq:rev_scaling} in two paradigmatic TFIMs that can be simulated efficiently at large system sizes: the uniform chain, \(J_i=1\), and the bond-disordered chain, with \(J_i\) drawn uniformly from \((0,2)\). The uniform model factorizes into independent Landau--Zener crossings, providing a benchmark for quasifree-fermionic systems. The disordered chain is integrable and has an infinite-randomness fixed point with \(z=\infty\), making it a particularly interesting test of the gKZM and our extension \cite{fisher_critical_1995,young_numerical_1996}. In this model, conventional KZM power-law scaling is expected only for fast quenches, while logarithmic corrections dominate at longer quench times. The model also captures physics relevant to systems featuring a random-singlet phase \cite{dasgupta_low-temperature_1980,igloi_strong_2018}.

For the forward quench protocol, the system is initialized in the ground state at \(h=0\) and ramped across the critical point to \(h=h_0=10\), allowing the defect density to be compared with the magnetization. The reverse quench protocol is initialized in the ground state at \(h=h_0=10\) and follows the ramp in Eq. \ref{eq:reverse_protocol}. We use system sizes \(N=10^4\) for the uniform TFIM and \(N=128\) for the bond-disordered TFIM. The dynamics are obtained by solving the time-dependent Bogoliubov--de Gennes equations \cite{dziarmaga_dynamics_2005,dziarmaga_dynamics_2006}. The expressions for the new observables, the defect density variance and the reverse quench defect density, are lengthy and are provided in the Supplemental Material \cite{supplement}.

The results are shown in Fig. \ref{fig:defects}. In the uniform TFIM, \(n_\mathrm{def}\), \(n_\mathrm{def}^\mathrm{rev}\), and \(\mathrm{Var}(n_\mathrm{def})\) all follow the predicted \(\tau_Q^{-1/2}\) scaling for \(\tau_Q > 10\). We also test Eqs. \ref{eq:gKZM} and \ref{eq:rev_scaling} by comparing \(n_\mathrm{def}/\mathrm{Var}(n_\mathrm{def}) = 1/(1-p) = 1/\left(1-1/\sqrt{2} \right) \), \(n_\mathrm{def}^\mathrm{rev}/\mathrm{Var}(n_\mathrm{def}) = 2\) and \(n_\mathrm{def}^\mathrm{rev}/n_\mathrm{def} = 2(1-p)=2-\sqrt{2}\). The inset shows agreement with both predictions within the error bars, supporting the validity of the gKZM and its extension to reverse quenches.

For the bond-disordered TFIM, \(n_\mathrm{def}\), \(n_\mathrm{def}^\mathrm{rev}\), and \(\mathrm{Var}(n_\mathrm{def})\) follow an approximate \(\tau_Q^{-1/2}\) scaling for \(2 \leq \tau_Q \lesssim 20\), characteristic of the uniform TFIM regime. In this fast-quench window, the relevant length scales are too short for disorder to significantly affect the dynamics \cite{dziarmaga_dynamics_2006}. At longer quench times, the observables depart from this power-law, consistent with disorder-dominated dynamics and previously reported deviations from power-law scaling of \(n_\mathrm{def}\) \cite{dziarmaga_dynamics_2006}.

For \(30 \leq \tau_Q \leq 140\), the three ratios \(n_\mathrm{def}/\mathrm{Var}(n_\mathrm{def})\), \(n_\mathrm{def}^\mathrm{rev}/n_\mathrm{def}\), and \(n_\mathrm{def}^\mathrm{rev}/\mathrm{Var}(n_\mathrm{def})\) are consistent with the gKZM predictions \(1/(1-p)\), \(2(1-p)\), and \(2\), respectively, for \(p \approx 0.37\). Thus, these ratio relations remain valid in a regime where the observables no longer follow the uniform-system power-law with \(\mu=1/2\), but instead reflect the disorder-dominated, \(\mu=0\) fixed point with logarithmic corrections.

At still longer quench times, \(n_\mathrm{def}^\mathrm{rev}/n_\mathrm{def}\) remains approximately constant even as the other two ratios depart from their gKZM values, hinting at their greater sensitivity to finite-size effects for this model. This suggests that a constancy of all three ratios, together with the condition \(n_\mathrm{def}^\mathrm{rev}/\mathrm{Var}(n_\mathrm{def})=2\), may provide an experimentally accessible indicator of the thermodynamic-limit scaling regime. A finite-size analysis of the uniform TFIM yields a similar picture (see Supplemental Material \cite{supplement}).

To further test the probabilistic description of defect formation, we extract \(p\) from two different ratios: the reverse quench defect density and the forward quench defect variance, each divided by the forward quench defect density. Using Eqs. \ref{eq:gKZM} and \ref{eq:rev_scaling}, these give
\begin{equation}
p = 1 - \frac{n^\mathrm{rev}_\mathrm{def}}{2 n_\mathrm{def}},
\label{eq:p_from_rev}
\end{equation}
and
\begin{equation}
p = 1 - \frac{\mathrm{Var}\left(n_\mathrm{def}\right)}{n_\mathrm{def}}.
\label{eq:p_from_var}
\end{equation}
If the gKZM and its extension hold, both estimates should agree and be independent of \(\tau_Q\).

\begin{figure*}[t]
    \centering
    \includegraphics[width=\linewidth]{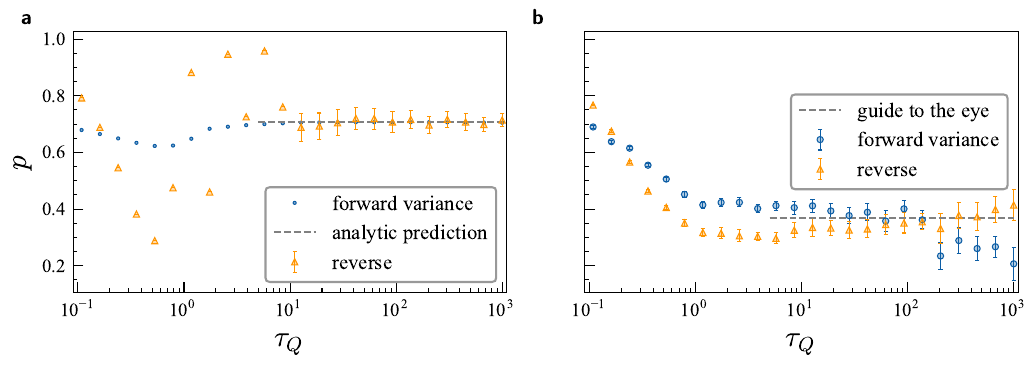}
    \caption{Estimated domain wall formation probability \(p\) in the gKZM as a function of the quench time \(\tau_Q\) for the uniform (a) and bond-disordered (b) TFIM. The probability is estimated both from a comparison between forward quench defect density and forward quench variance (blue dots, Eq. \ref{eq:p_from_var}) as well as from a comparison between the reverse quench defect density and the forward quench defect density (orange triangles, Eq. \ref{eq:p_from_rev}). In the latter case, error bars are propagated from the reverse quench defect density. The gray dashed line shows the analytic prediction \(p=1/\sqrt{2}\) in the uniform case, and \(p=0.37\), cf. Fig. \ref{fig:defects}.}
    \label{fig:probs}
\end{figure*}

The results are shown in Fig. \ref{fig:probs}. For the uniform TFIM, both estimates of \(p\) agree with the predicted value \(1/\sqrt{2}\) for \(\tau_Q > 10\), with the reverse quench estimate averaged over approximately two Stückelberg oscillations. In the bond-disordered TFIM, the two estimates agree for \(30 \leq \tau_Q \leq 140\). This is precisely the regime in which the ratio of the forward quench defect density to its variance is independent of \(\tau_Q\), and the reverse quench defect density is twice the forward quench variance, as predicted by the gKZM and our extension. Thus, within this window, defect formation is consistent with independent stochastic events characterized by a \(\tau_Q\)-independent probability \(p\), and the defect cumulant scaling is dominated by the \(\mu=0\) exponent.

At larger \(\tau_Q\), the agreement between the ratios and their predicted values gradually deteriorates. For quench times of order \(10^2\), the expected number of defects in the \(N=128\) system approaches unity, consistent with a frozen correlation length comparable to the system size. Finite-size effects thus grow increasingly important as \(\tau_Q\) increases, and the dynamics begin to cross over toward the adiabatic regime, obscuring the thermodynamic-limit gKZM scaling. This is consistent with previous findings that, when the number of defects in the system is \(\mathcal{O}(1)\), a crossover to adiabatic dynamics can alter the scaling \cite{dziarmaga_dynamics_2005,caneva_adiabatic_2007,schmitt_quantum_2022}, and can also be observed in the uniform TFIM for smaller system sizes and larger quench times (see Supplemental Material \cite{supplement}).

For both models considered here, and more generally for systems with a paramagnetic phase, the reverse quench defect density can be estimated at sufficiently large magnetic fields as
\begin{equation}
    n_\mathrm{def} \approx \frac{1}{2} - \langle \hat{S}_x \rangle,
\end{equation}
where \(\hat{S}_x\)  is the total magnetization along the applied field, normalized per spin, so that a fully polarized state has magnetization \(1/2\)  \cite{dziarmaga_dynamics_2005}. Thus, if the reverse quench begins and ends at a sufficiently large field, the defect density can be inferred from a single global magnetization measurement at the end of the protocol, without resolving individual defects, while performing the reverse quench requires only global magnetic field control. Thus, quantifiying critical scaling is possible relying on global system access only (see detailed simulations of the magnetization in the Supplementary Material \cite{supplement}).

In conclusion, we establish that, within the regime of validity of the stochastic gKZM, the mean defect density after a reverse quench equals twice the forward quench defect-number variance per unit volume. Agreement with this relation in uniform and bond-disordered TFIMs, including a regime without power-law defect scaling, shows that it applies beyond the uniform-system scaling regime. The result also suggests an experimental route to probing critical dynamics: global field control and magnetization measurements can reveal reverse quench defect densities without resolving individual defects. This approach may be useful in quantum spin glasses and other systems with challenging critical dynamics \cite{harris_phase_2018,king_quantum_2023,kroeze_directly_2025}.

\begin{acknowledgments}
We thank T. Gasenzer for fruitful discussions. E. J. Braun acknowledges support by the IMPRS for Quantum Dynamics in Physics, Chemistry and Biology. This work is part of and supported by the Deutsche Forschungsgemeinschaft (DFG, German Research Foundation) under Germany’s Excellence Strategy EXC2181/1-390900948 (the Heidelberg STRUCTURES Excellence Cluster), within the Collaborative Research Centre “SFB 1225 (ISOQUANT)”, the DFG Priority Program “GiRyd 1929”, the Horizon Europe programme HORIZON-CL4-2022-QUANTUM-02-SGA via the project 101113690 (PASQuanS2.1), and the Heidelberg Center for Quantum Dynamics. The authors acknowledge support by the state of Baden-Württemberg through bwHPC and the German Research Foundation (DFG) through grant INST 35/1597-1 FUGG (Helix cluster).
\end{acknowledgments}
\appendix

\bibliographystyle{apsrev4-2}
\bibliography{bib}

\end{document}


\section{Detailed plots of cumulant ratios}

\begin{figure*}[htb]
    \centering
    \includegraphics[width=\linewidth]{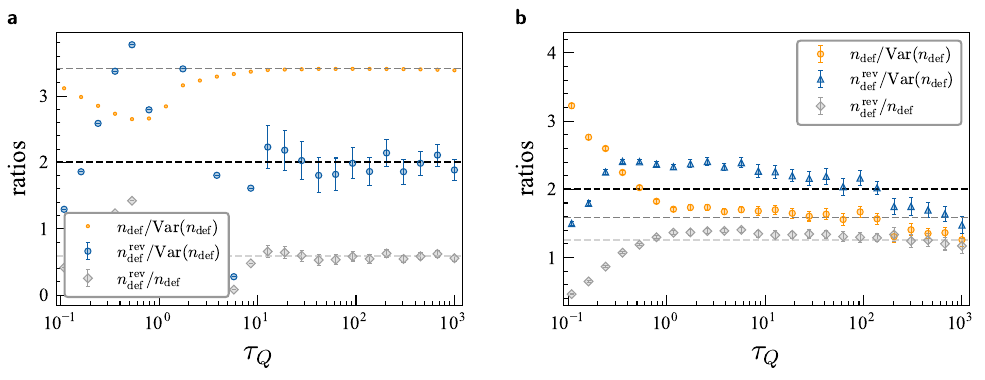}
    \caption{Ratios of the forward and reverse quench defect densities to the forward quench defect density variance (orange and blue, respectively) and ratio between reverse and forward quench defect density (gray) for the uniform (left panel) and bond-disordered (right panel) TFIM. The black, gray and light gray dashed lines indicate \(2\), the approximate value \(1/(1-p)\) and \(2(1-p\), where \(p=\frac{1}{\sqrt{2}}\) in the uniform and \(p=0.37\) in the bond-disordered TFIM, respectively.}
    \label{fig:ratios}
\end{figure*}

In Fig. \ref{fig:ratios}, we show a magnified view of the cumulant ratios as functions of quench time, corresponding to the insets of Fig. 2 in the main text. 

For the uniform model, all three discussed cumulant ratios agree with their predicted values for \(\tau_Q \ge 10\). Notably, the regime where the average over Stückelberg oscillations start becoming well defined, is the same where the ratio \(n_\mathrm{def}/\mathrm{Var}(n_\mathrm{def})\) becomes constant and consistent with the analytic prediction of \(1/(1-1/\sqrt{2})\approx 3.41\). As a consequence, the constancy of all three ratios, together with the condition \(n_\mathrm{def}^\mathrm{rev}/\mathrm{Var}(n_\mathrm{def})=2\), may serve as an indicator that the universal gKZM scaling regime is achieved and transient model-specific initial behavior does not govern the scaling of defects and their cumulants anymore.

For the bond-disordered model, we observe that already for \(2 \le \tau_Q \le 20\), all three ratios reach a value independent of \(\tau_Q\). However, in this regime, as is visible in Fig. 2 in the main text, the scaling is still governed by the uniform model fixed point and critical exponents, so the regime is transient. This becomes most clearly visible, as the ratio \(n_\mathrm{def}^\mathrm{rev}/\mathrm{Var}(n_\mathrm{def})\) differs significantly from 2, by more than a\(3\sigma\) confidence interval. The only values that agree within a \(1\sigma\) confidence interval are \(\tau_Q\approx 60\) and \(\tau_Q \approx 140\). Within a \(2\sigma\) confidence interval, the ratio \(n_\mathrm{def}^\mathrm{rev}/\mathrm{Var}(n_\mathrm{def})=2\) only in the range \(30 \le \tau_Q \le 140\), which is also the range where the defect formation probability is both independent of \(\tau_Q\) and independent of the method it is derived (cf. Fig. 3 in the main text). For later \(\tau_Q\), where finite-size effects are expected to derive, all ratios are again \(\tau_Q\)-dependent.

Thus, this numerical study indicates that the constancy of all three ratios, together with the condition \(n_\mathrm{def}^\mathrm{rev}/\mathrm{Var}(n_\mathrm{def})=2\), might be an experimental signature of which window of quench times corresponds to the actual critical scaling regime, and is not dominated neither by transient dynamics nor by finite size effects. We discuss this point further in Sec. \ref{sec:finite_size} on different size-dependent effects for the uniform TFIM.

\section{Analytic derivation of the uniform TFIM defects}

\subsection{Bogoliubov-de Gennes equations in the uniform TFIM}
This section mainly follows \cite{dziarmaga_dynamics_2005}. As a starting point, we use the Hamiltonian of the uniform TFIM with an even number of spins \(N\) and periodic boundary conditions \(\sigma_{N+1}=\sigma_1\)
\begin{equation}
    H = -\sum_{i=1}^N J_i \sigma_i^z \sigma_{i+1}^z -h \sum_{i=1}^N \sigma_i^x.
    \label{eq:H_Ising}
\end{equation}
Applying a Jordan-Wigner transform to fermionic operators
\begin{align}
    \label{eq:sigmaz}
    \sigma_i^z &= -\prod_{j<i}(1-2c_j^\dagger c_j)(c_i+c_i^\dagger),
    \\
    \label{eq:sigmax}
    \sigma_i^x &= 1-2c_i^\dagger c_i,
\end{align}
we obtain the spinless fermionic Hamiltonian 
\begin{equation}
    H = \sum_{i=1}^N  J_i(c_i-c_i^\dagger)(c_{i+1}+c_{i+1}^\dagger)+h(2c_i^\dagger c_i-1),
    \label{eq:Ham_ferm}
\end{equation}
where the periodic boundaries of the spin Hamiltonian translate to anti-periodic boundary conditions on the fermionic operators \(c_{N+1}=-c_1\) if we assume dynamics only in the positive parity subspace, to which both the ground state and the fully spin polarized belong to \cite{he_boundary_2017}.
For the uniform TFIM where \(J_i=1\) for all \(i\), we can apply a Fourier transform to modes labeled by their wavevector \(k \in \left\{\pm \frac{\pi}{N},\pm \frac{3\pi}{N},\dots,\pm \frac{(N-1)\pi}{N}\right\}\) to obtain
\begin{equation}
    H = \sum_{k=-\frac{N-1}{N\pi}}^{\frac{N-1}{N\pi}} 2(h-\cos(k))c_k^\dagger c_k + \sin(k) (c_k^\dagger c_{-k}^\dagger + c_{-k}c_k) -h,
    \label{eq:H_FT}
\end{equation}
where only modes \(\pm k\) interact with one another.

To obtain the ground state of the system, and finally solve the dynamics, we seek to bring this Hamiltonian to the form
\begin{equation}
    H = \sum_k \epsilon_k \left(\gamma_k^\dagger \gamma_k-\frac{1}{2}\right),
    \label{eq:qp}
\end{equation}
where \(\gamma_k\) is the annihilation operator for a fermionic quasiparticle, and \(\epsilon_k\) the energy associated with it. This can be achieved by applying the Bogoliubov transformation
\begin{equation}
    c_k = u_k \gamma_k + v_{-k}^\ast \gamma_{-k}^\dagger,
    \label{eq:Bog_transf}
\end{equation}
where the prefactors \(u_k\) and \(v_k\) fulfill
\begin{equation}
    \epsilon_k \begin{pmatrix}
        u_k \\ v_k
    \end{pmatrix} = \begin{pmatrix}
        2(h-\cos(k)) & 2\sin(k) \\
        2\sin(k) & -2(h-\cos(k))
    \end{pmatrix} \begin{pmatrix}
        u_k \\ v_k
    \end{pmatrix}.
    \label{eq:BdG_energies}
\end{equation}

This equation has a positive and a negative solution with energy \(\pm \epsilon_k\). The positive-energy eigenvector is associated with the creation operator of the quasiparticle \(\gamma_k^\dagger\), whereas the negative-energy solution is associated with the creation operator of a hole, equivalent to the annihilation of a quasiparticle \(\gamma_{-k}\). The ground state is thus the state annihilated by all \(\gamma_k\).

To derive the dynamics in this Hamiltonian, we switch to the Heisenberg picture where operators evolve in time and states are time-independent. As the state, we choose the quasiparticle vacuum \(\ket{0}\) which is defined by \(\gamma_k \ket{0}=0 \) for all \(k\). In this picture, the time-independent state is the ground state of the Hamiltonian at the initial time, i.e. the quasiparticle vacuum which is annihilated by all \(\gamma_k\) that are positive-energy solution of Eq. \ref{eq:BdG_energies}. In the limit \(h \rightarrow \infty\), the ground state is the fully polarized state \(\ket{\rightarrow}^{\otimes N}\), which is annihilated by \(\gamma_k\) with \(\left(u_k,v_k\right) = \left(1,0\right) \).

The time evolution of the fermionic operators can be calculated using the Heisenberg equation of motion
\begin{equation}
    i \frac{dc_k}{dt} = \comm{c_k}{H}
    \label{eq:Heisenberg_ck}
\end{equation}
together with the constraint \(\frac{d\gamma_k}{dt}=0\). Applying Eq. \ref{eq:Bog_transf} and thus taking all the time-dependence into the \(u_k\) and \(v_k\), we obtain the time-dependent Bogoliubov-de Gennes equations
\begin{widetext}
\begin{equation}
    i\frac{d}{dt} \begin{pmatrix}
        u_k(t) \\ v_k(t)
    \end{pmatrix} = \begin{pmatrix}
        2(h(t)-\cos(k)) & 2\sin(k) \\
        2\sin(k) & -2(h(t)-\cos(k))
    \end{pmatrix} \begin{pmatrix}        u_k(t) \\ v_k(t)
    \end{pmatrix}.
    \label{eq:tdBdG}
\end{equation}
\end{widetext}
\subsection{Analysis of forward and reverse quench defect densities}

As detailed in the main text, we consider a time-dependent magnetic field
\begin{equation}
    h(t) = \frac{\left|t\right|}{\tau_Q},
\end{equation}
where \(t\) runs from \(-\infty\) to zero for a forward quench, and further up to \(\infty\) for a reverse quench. This is equivalent to letting \(t\) run from \(-\infty\) to \(0\), and then back to \(-\infty\), which allows us to rewrite Eq. \ref{eq:tdBdG} as

\begin{widetext}
\begin{equation}
    i\frac{d}{dt} \begin{pmatrix}
        u_k(t) \\ v_k(t)
    \end{pmatrix}
    = \begin{pmatrix}
        2(-\frac{t}{\tau_Q}-\cos(k)) & 2\sin(k) \\
        2\sin(k) & -2(-\frac{t}{\tau_Q}-\cos(k))
    \end{pmatrix} \begin{pmatrix}        u_k(t) \\ v_k(t)
    \end{pmatrix}.
    \label{eq:tdBdG2}
\end{equation}
\end{widetext}
Applying the substitution
\begin{equation}
    \tau = 4\tau_Q\sin(k) \left(\frac{t}{\tau_Q}+\cos(k)\right),
    \label{eq:tau_of_t}
\end{equation} Eq. \ref{eq:tdBdG2} can be brought into the form
\begin{equation}
    i\frac{d}{d\tau} \begin{pmatrix}
        u_k(\tau) \\ v_k(\tau)
    \end{pmatrix} = \begin{pmatrix}
        -1/2 \tau \Delta_k & 1/2 \\
        1/2 & 1/2 \tau \Delta_k 
    \end{pmatrix} \begin{pmatrix}
        u_k(\tau) \\ v_k(\tau)
    \end{pmatrix},
    \label{eq:LZ_form}
\end{equation}

which features the Hamiltonian of Landau-Zener crossing with gap \(\Delta_k = \left(4\tau_Q\sin^2(k)\right)^{-1}\).
According to Eq. \ref{eq:tau_of_t}, \(\tau\) thus runs from \(-\infty\) to \(\tau_\mathrm{final}=2\tau_Q\sin(2k)\) in the forward quench, and back to \(-\infty\) in the reverse quench. We observe that
\begin{equation}
    \tau_\mathrm{final}\Delta_k = \frac{1}{2} \cot(k) \gg1 , \quad k\ll \pi,
\end{equation}
i.e. that as long as only long-wavelength modes are excited, the final time is far outside the region where tunneling between ground and excited state are possible in the Landau-Zener passage, and thus for these long-wavelength modes, the quench can be extended for \(\tau\) running from negative to positive infinity for the forward quench, and back for the reverse quench.

For each such mode \(k\), the excitation probability is
\begin{equation}
    p_k = e^{-\frac{\pi}{2\Delta_k}}.
    \label{eq:transition_prob}
\end{equation}
For slow quenches where only the long wavelength modes get excited, such that \(p_k\) is strongly concentrated only near \(k=0\), and can be approximated for \(k \approx 0\) as a Gaussian
\begin{equation}
    p_k \approx e^{-2 \pi \tau_Q k^2}.
    \label{eq:transition_prob_approx}
\end{equation}
In the thermodynamic limit \(N\rightarrow \infty\), the defect density is thus given by
\begin{equation}
\begin{split}
   n_\mathrm{def} &= \lim_{N\rightarrow \infty}\frac{1}{N}\sum_k p_k = \frac{1}{2\pi}\int_{-\pi}^\pi p_k dk\\
   &\approx \frac{1}{2\pi}\int_{-\infty}^\infty p_k dk = \frac{1}{2\pi} \frac{1}{\sqrt{2\tau_Q}}\sim \tau_Q^{-\frac{1}{2}}.
   \end{split}
    \label{eq:analytic_ndef}
\end{equation}
This approximation is valid as long as standard deviation of the Gaussian distribution is small compared to \(2\pi\). As \(p_k\) is a Gaussian function centered around 0, if the standard deviation of the Gaussian function is small compared to \(2\pi\), extending the integration boundaries from \(\left(-\pi,\pi\right)\) to \(\left(-\infty,\infty\right)\) leaves the integral approximately unchanged. This condition is fulfilled for \(\frac{1}{\sqrt{4\pi \tau_Q}} \ll \pi\), which is true for long quench times, and self-consistently shows that only long-wavelength modes become excited.

A detailed derivation of the defect density variance after a forward quench from the moment generating functional is given in \cite{del_campo_universal_2018,cui_experimentally_2020}. Here, we choose a different derivation, using Eq. \ref{eq:H_FT} as a starting point. Because the whole system separates into \(N\) independent subsystems consisting of modes \(\pm k\), we can calculate higher order cumulants of the defect density by calculating cumulants in the individual subsystems. As a consequence, the defect density variance is the average of the excitation variance in each single subsystem. Therefore, we can omit the calculation of the moment generating functional of the defects in the entire system. Instead, we treat each subsystem labeled by its mode \(k\) individually, which comprises a 2-level system undergoing a Landau-Zener transition. The subsystem is initialized in its ground state \(\ket{g}\) and undergoes a time evolution to its final state \(\ket{g(t_\mathrm{final})}\). The number operator that counts the amount of excitations is represented by the operator \(\hat{n}=\dyad{e}\). As stated in Eq. \ref{eq:transition_prob}, according to the Landau-Zener-formula, the expectation value of the number operator equals the excitation probability
\begin{equation}
    \ev{\hat{n}}{g(t)}=p_k = \left|\ip{g(t)}{e}\right|^2.
\end{equation}
In addition, since \(\dyad{e}\) is a projection operator, it fulfills 
\begin{equation}
    \hat{n}^m = \dyad{e}^m = \dyad{e} = \hat{n}
\end{equation} for all positive integer values \(m\). As a consequence, the expectation value of higher powers of the number operator is equal to the excitation probability
\begin{equation}
    \ev{\hat{n}^m}{g(t)}= \ev{\hat{n}}{g(t)} = p_k
\end{equation}
for every \(m\). As a consequence, the number variance for a single mode \(k\) is given by
\begin{equation}
   \mathrm{Var}(n_\mathrm{def}(k))= \expval{n_\mathrm{def}(k)^2}-\expval{n_\mathrm{def}(k)}^2=p_k-p_k^2.
\end{equation}
In the continuum limit and in the limit of large quench times such that only long wavelength modes get excited, the variance of the number of defects in the system normalized by the system size is thus given by
\begin{equation}
\begin{split}
    \mathrm{Var}(n_\mathrm{def}) &= \frac{1}{2\pi} \int_{-\pi}^\pi \left(p_k - p_k^2\right) dk\\
    & \overset{\ref{eq:transition_prob}}{=} \frac{1}{2\pi} \int_{-\pi}^\pi \left(e^{-\frac{\pi}{2 \Delta_k}} - e^{-\frac{\pi}{\Delta_k}}\right) dk\\
    & \overset{\ref{eq:transition_prob_approx}}{=} \frac{1}{2\pi} \int_{-\infty}^\infty \left(e^{-2\pi \tau_Q k^2}-e^{-4\pi \tau_Q k^2}\right) dk\\
    &= \frac{1}{2\pi} \left(\frac{1}{\sqrt{2\tau_Q}} - \frac{1}{\sqrt{4\tau_Q}} \right)\\
    &= \left(1-\frac{1}{\sqrt{2}}\right) n_\mathrm{def}.
\end{split}
\label{eq:analytic_var_ndef}
\end{equation}
Again, we assumed that only long wavelength modes become occupied, which is valid for \(\frac{1}{\sqrt{4\pi \tau_Q}}, \frac{1}{\sqrt{8\pi\tau_Q}} \ll \pi\), i.e. for large \(\tau_Q\).

For the uniform TFIM, the reverse quench protocol was solved exactly for arbitrary quench times in \cite{kou_interferometry_2022}. A heuristic approach neglecting interference effects between defects that were created in the first quench and defects created in the second quench, i.e. Stückelberg oscillations \cite{stueckelberg_theorie_1932,shevchenko_landauzenerstuckelberg_2010,kou_interferometry_2022,ivakhnenko_nonadiabatic_2023}, can be found in \cite{mukherjee_defect_2008}. Here, we choose a different approach incorporating the Stückelberg oscillations, approximating the defect density for large quench times.

For the treatment of the Stückelberg oscillations, we follow mainly \cite{ivakhnenko_nonadiabatic_2023}.
For a single \(k\)-mode, the reverse quench protocol is equivalent to a double passage of a Landau-Zener crossing. In this double passage, an interference effect occurs when measuring the excitation probability. The path in which an excitation occurs in the first crossing but not in the second one interferes with the path in which an excitation occurs in the second crossing but not in the first one. As a consequence, the excitation probability after a double passage is given by
\begin{equation}
    p_k (\tau_Q) = 4 e^{-\frac{\pi}{2\Delta_k}}\left(1-e^{-\frac{\pi}{2\Delta_k}}\right)\sin^2(\Phi_\mathrm{St}),
    \label{eq:Stueckelberg}
\end{equation}

where the phase difference between the two excitation paths is the Stückelberg phase \(\Phi_\mathrm{St}\), which is given by
\begin{equation}
    \Phi_\mathrm{St} = \zeta_1 +\phi_S.
\end{equation} 
\(\zeta_1\) is the phase acquired during an adiabatic evolution and \(\phi_S\) is the Stokes phase which is acquired due to the non-adiabatic nature of a forward quench. \(\zeta_1\) can be calculated from the energy difference between the two Landau-Zener crossings, i.e.
\begin{equation}
\begin{split}
    \zeta_1(k) &= \frac{1}{2} \int_{t_1}^{t_2} \Delta E_k(t) dt\\
    &= 2 \int_{t_1}^{t_2} \sqrt{\left(\frac{t}{\tau_Q}\right)^2 - 2 \frac{\left|t\right|}{\tau_Q} \cos(k)+1}dt,
\end{split}
\end{equation}
where \(t_1=-\tau_Q\cos(k)\) and \(t_2=\tau_Q\cos(k)\) are the times at which the gap is minimal. For small \(k \ll \frac{\pi}{4}\) which are excited for long enough quench times, \(\cos(k) \approx 1\). Making the substitution \(u=\frac{t}{\tau_Q}\), the phase \(\zeta_1\) can be approximated by
\begin{equation}
    \zeta_1(k) \approx 2 \tau_Q \int_{-1}^1 \sqrt{u^2-2\left|u\right|+1} du = 2\tau_Q,
\end{equation}
which for small \(k\) is nearly \(k\)-independent. On the other hand, for large quench times, the Stokes phase \(\phi_S\) is known to be approximately zero \cite{ivakhnenko_nonadiabatic_2023}. As such, the Stückelberg phase for large quench times is nearly \(2\tau_Q\) \cite{ivakhnenko_nonadiabatic_2023,shevchenko_landauzenerstuckelberg_2010}, leading to an oscillation period of \(\pi/2\). For large quench times, the excitation probability can be approximated as a constant during a Stückelberg oscillation. Therefore, the excitation probability can be treated as a slowly evolving envelope, on which fast Stückelberg oscillations occur. Thus, averaging only the Stückelberg oscillation over one oscillation period yields the average excitation probability \(\overline{p_k}\)

\begin{equation}
\begin{split}
    \overline{p_k} &= 4 e^{-\frac{\pi}{2\Delta_k}}\left(1-e^{-\frac{\pi}{2\Delta_k}}\right) \frac{2}{\pi} \int_0^{\frac{\pi}{2}} \sin^2(2\tau_Q) d\tau_Q\\
    &= 2 e^{-\frac{\pi}{2\Delta_k}}\left(1-e^{-\frac{\pi}{2\Delta_k}}\right)
    \end{split}
\label{eq:prob_with_Stuckelberg}
\end{equation}

in the limit \(\tau_Q \gg \pi/2\). We use this to calculate the defect density after the reverse quench \( \overline{n_\mathrm{def}^\mathrm{rev}} \) averaged over one Stückelberg oscillation period. In the limit of large \( \tau_Q \), we assume self-consistently that only long-wavelength modes are excited. The average reverse quench defect density then reads

\begin{equation}
\begin{split}
    \overline{n_\mathrm{def}^\mathrm{rev}} &= \frac{1}{2\pi} \int_{-\pi}^\pi \overline{p_k} dk\\
    &\approx \frac{1}{2\pi} \int_{-\pi}^\pi 2 e^{-\frac{\pi}{2\Delta_k}}\left(1-e^{-\frac{\pi}{2\Delta_k}}\right) dk\\
    &\approx \frac{1}{2\pi} \int_{-\infty}^\infty 2 e^{-2\pi \tau_Q k^2}\left(1-e^{-2\pi \tau_Q k^2}\right) dk\\
    &= \frac{1}{2\pi} 2 \left(\frac{1}{\sqrt{2\tau_Q}} -  \frac{1}{\sqrt{4\tau_Q}}\right)\\
    &= \left(2-\sqrt{2}\right) \frac{1}{2\pi} \frac{1}{\sqrt{2 \tau_Q}} = \left(2-\sqrt{2}\right) n_\mathrm{def}.
\end{split}
\label{eq:analytic_revdef}
\end{equation}

The assumption that only long wavelength modes are excited is valid for \(\frac{1}{\sqrt{4\pi \tau_Q}}, \frac{1}{\sqrt{8\pi\tau_Q}} \ll \pi\), which is valid for large \(\tau_Q\).

\section{Numerical implementation of the uniform TFIM}

We consider a uniform TFIM with \(N=10^4\) spins and nearest-neighbor coupling \(J=1\) with periodic boundary conditions. The transverse magnetic field \(h\) undergoes the time evolution 
\begin{equation}
    h(t)=\frac{t}{\tau_Q}, \quad 0 \le t \le h_0 \tau_Q
    \label{eq:field_numerics_forward}
\end{equation}
or
\begin{equation}
    h(t)=\frac{\left|t\right|}{\tau_Q}, \quad -h_0 \tau_Q \le t \le h_0\tau_Q
    \label{eq:field_numerics_reverse}
\end{equation}
for the forward and reverse quench protocols, respectively, where \(h_0=10\). We vary the quench time \(\tau_Q\) between \(10^{-2} \le \tau_Q \le 10^3 \), choosing 30 logarithmically spaced values of \(\tau_Q\) between \(10^{-2}\) and \(10^{3}\). Unlike in the analytical model, we the forward quench starts here in the ground state of the ferromagnetic phase. However, as shown in Ref. \cite{dziarmaga_dynamics_2005}, the direction of the forward quench does not affect the number of created defects as a function of quench duration.

The system is initialized in its ground state at initial time \(t_\mathrm{init}\). For efficient calculation, we use Eq. \ref{eq:BdG_energies} in order to get the initial values for both the negative-energy eigenvector \(u_k^-(t_\mathrm{init})\) and \(v_k^-(t_\mathrm{init})\) as well as for the positive-energy eigenvector \(u_k(t_\mathrm{init})\) and \(v_k(t_\mathrm{init})\). We use the values of the positive-energy eigenvector to initialize the system in the ground state, together with Eq. \ref{eq:tdBdG}, to obtain the final values \(u_k(h_0 \tau_Q) \coloneqq \tilde{u}_k\) and \(v_k(h_0 \tau_Q) \coloneqq \tilde{v}_k\). In addition, we use Eq. \ref{eq:BdG_energies} to get the negative eigenmode \(\bar{u}_k\) and \(\bar{v}_k\), corresponding to the eigenmode of a quasiparticle excitation at the final field \(h_0=10\). In order to solve the differential equation \ref{eq:tdBdG}, we make use of the DifferentialEquations.jl package \cite{rackauckas2017differentialequations}. The defect density is then given by total number of quasiparticles, which using Eq. \ref{eq:Bog_transf} corresponds to an overlap between the final states \(\left(\tilde{u}_k, \tilde{v_k}\right)\) and the negative eigenstates \(\left(\bar{u}_k,\bar{v}_k\right)\)
\begin{equation}
\begin{split}
    n_\mathrm{def} &= \frac{1}{N} \sum_k \left| \begin{pmatrix}
      \bar{u}_k^\ast &  \bar{v}_k^\ast  
    \end{pmatrix} \begin{pmatrix}
      \tilde{u}_k \\  \tilde{v_k}
    \end{pmatrix}\right|^2\\
    &= \frac{1}{N} \sum_k \begin{pmatrix}
      \tilde{u}_k^\ast &  \tilde{v_k}^\ast
    \end{pmatrix} \begin{pmatrix}
        \bar{u}_k \bar{u}_k^\ast & \bar{u}_k \bar{v}_k^\ast\\
        \bar{v}_k \bar{u}_k^\ast & \bar{v}_k \bar{v}_k^\ast
    \end{pmatrix}
    \begin{pmatrix}
        \tilde{u}_k \\ \tilde{v}_k
    \end{pmatrix}\\
    &= \frac{1}{N} \sum_k \begin{pmatrix}
      \tilde{u}_k^\ast &  \tilde{v_k}^\ast
    \end{pmatrix} \hat{O}_k
    \begin{pmatrix}
        \tilde{u}_k \\ \tilde{v}_k
    \end{pmatrix}\\
    &= \frac{1}{N} \sum_k p_k,
\end{split}
    \label{eq:def_ordered}
\end{equation}
where \(\hat{O}_k\) represents the operator that measures the defect number in mode \(k\) at the field \(h_0\), and \(p_k\) thus corresponds to the probability that a defect was created in mode \(k\). For the implementation, we calculate the overlap between the final states and the negative-energy eigenstates.

To calculate the defect density variance, we use the fact that each \(\pm k\)-subsystem as defined in Eq. \ref{eq:tdBdG} is independent of the other subsystems with different wavevectors. As such, the variance of the whole system is the sum of the variances of the individual subsystems. Thus, the variance of the defect density can be calculated from the variance of the operator \(\hat{O}_k\). Note that due to the fermionic nature of both \(c_k\) and \(\gamma_k\), \(\left|\bar{u}_k\right|^2+\left|\bar{v}_k\right|^2=1\), which leads to \(\hat{O}_k^2 = \hat{O}_k\). Thus, the defect variance is given by
\begin{equation}
\begin{split}
    &\mathrm{Var}(n_\mathrm{def})\\
    & = \frac{1}{N} \sum_k \left[ \begin{pmatrix}
      \tilde{u}_k^\ast &  \tilde{v_k}^\ast
    \end{pmatrix} \hat{O}_k^2
    \begin{pmatrix}
        \tilde{u}_k \\ \tilde{v}_k
    \end{pmatrix}
    - \left( \begin{pmatrix}
      \tilde{u}_k^\ast &  \tilde{v_k}^\ast
    \end{pmatrix} \hat{O}_k
    \begin{pmatrix}
        \tilde{u}_k \\ \tilde{v}_k
    \end{pmatrix}\right)^2\right]\\
    &= \frac{1}{N} \sum_k \left(p_k - p_k^2\right).
\end{split}
    \label{eq:var_ordered}
\end{equation}

Deep in the paramagnetic regime, when the external field \(h = \infty\), which is equivalent to \(J=0\), i.e. only the interaction with an external field is present, the defect number operator can be immediately related to the magnetization operator
\begin{equation}
    \hat{S}_x = \frac{1}{2N} \sum_i \sigma_i^x = \frac{1}{2} - \hat{n}_\mathrm{def}.
    \label{eq:mag_def}
\end{equation}
To show this relation, we write the Hamiltonian in the limit of infinite magnetic field after a Jordan-Wigner transform
\begin{equation}
\begin{split}
    H &= \sum_i h\left(2c_i^\dagger c_i -1 \right) = \sum_i 2h \left(c_i^\dagger c_i -\frac{1}{2}\right)\\
    & \overset{\mathrm{FT}}{=} \sum_k 2h \left(c_k^\dagger c_k -\frac{1}{2}\right).
\end{split}
\end{equation}
A comparison to Eq. \ref{eq:qp} shows that in the pure paramagnet, the equation \(\gamma_k = c_k\) holds, such that we obtain in the paramagnet
\begin{equation}
    H = \sum_k 2h \left(\gamma_k^\dagger \gamma_k -\frac{1}{2}\right) = -h \sum_i \sigma_i^x = -2h \sum_i \hat{S}_i^x.
\end{equation}
Using the definition of the defect density \(\hat{n}_\mathrm{def} = \sum_k \gamma_k^\dagger \gamma_k\) and of the magnetization operator in \(x\)-direction \(N\hat{S}_x = \sum_i \hat{S}_i^x\) and dividing the previous equation by \(-2hN\), we obtain
\begin{equation}
    -\hat{n}_\mathrm{def} + \frac{1}{2} = \hat{S}_x ,
\end{equation}
which proves Eq. \ref{eq:mag_def}. As a consequence, a measurement of magnetization of the system at a final field \(h\) is equivalent to an instantaneous quench of the system from this final value to \(h=\infty\) and a measurement of \(1/2-\hat{n}_\mathrm{def}\).

As in this limit \(c_k=\gamma_k\), the positive eigenmode of Eq. \ref{eq:BdG_energies} is given by \(\left(u_k,v_k\right)= \left(1,0\right)\) and the negative eigenmode by \(\left(u_k,v_k\right)= \left(0,1\right)\). As a consequence, we find for all modes \(k\) in this limit \(\bar{u}_k = 0 \land \bar{v}_k=1 \). Plugging this into Eq. \ref{eq:def_ordered}, we find for the expectation value of the magnetization in the states \(\left(\tilde{u}_k,\tilde{v}_k\right)\)

\begin{equation}
    \expval{\hat{S}_x} =\frac{1}{2}-\frac{1}{N}\sum_k \left|\tilde{v}_k\right|^2.
    \label{eq:mag_ordered}
\end{equation}
Starting from Eq. \ref{eq:mag_def}, we obtain for the averaged magnetization fluctuations in the pure paramagnet
\begin{equation}
\begin{split}
    \mathrm{Var}\left(\hat{S}_x\right) &= \frac{1}{N} \left(\expval{N^2 {\hat{S}_x}^2} - \expval{N \hat{S}_x}^2\right)\\
    &= N \left( \expval{\left(\frac{1}{2} - \hat{n}_\mathrm{def}\right)^2}  - \expval{\frac{1}{2} - \hat{n}_\mathrm{def}}^2\right)\\
    &= N \left( \frac{1}{4} - \expval{\hat{n}_\mathrm{def}} + \expval{\hat{n}_\mathrm{def}^2} - \frac{1}{4} + \expval{\hat{n}_\mathrm{def}} - \expval{n_\mathrm{def}}^2\right)\\
    &= N \left(\expval{\hat{n}_\mathrm{def}^2} - \expval{n_\mathrm{def}}^2 \right)\\
    &= \frac{1}{N} \left(\expval{\hat{N}_\mathrm{def}^2} - \expval{\hat{N}_\mathrm{def}}^2 \right) = \mathrm{Var}\left(n_\mathrm{def}\right).
\end{split}
\end{equation}

So a measurement of the magnetization fluctuation corresponds to an instantaneous quench of the system to the pure paramagnet and a measurement of the average defect fluctuations. Comparing to Eq. \ref{eq:var_ordered}, we find 
\begin{equation}
    \mathrm{Var}\left(\hat{S}_x\right) = \frac{1}{N} \sum_k \left|\tilde{v}_k \right|^2 - \left|\tilde{v}_k \right|^4.
    \label{eq:magvar_ordered}
\end{equation}

In the numerical simulation, after calculating the values \(\bar{u}_k\), \(\bar{v}_k\), \(\tilde{u}_k\) and \(\tilde{v}_k\), we use Eqs. \ref{eq:def_ordered} and \ref{eq:var_ordered} to obtain the defects and their variance, as well as Eqs. \ref{eq:mag_ordered} and \ref{eq:magvar_ordered} in order to get the magnetization and its variance. For the numerical integration of the differential equation, we use the TSit5 algorithm, implemented in \cite{rackauckas2017differentialequations}.

For the reverse quench defect density, we have to consider how to average over Stückelberg oscillations introduced in Eq. \ref{eq:Stueckelberg}. For \(\tau_Q < 10\), the Stückelberg oscillations with angular frequency 4 are not fast compared to the change of the trend of the excitation probability. As a consequence, we take no time-averages.

 For \(\tau_Q > 10\), the angular frequency of the Stückelberg oscillations is approximately 4 and their amplitude is approximately constant. This allows to average the defect density or the magnetization with 11 points linearly spaced in the interval \(\left[\tau_Q-2,\tau_Q+2 \right]\), roughly corresponding to two oscillation periods. The mean of these values is a numerical approximation of \(\overline{p_k}\), while the error of the mean, i.e. the standard deviation of these values divided by the square root of the number of data points, 11, is an estimate for the error bar of \(\overline{p_k}\).

\section{Numerical implementation of the bond-disordered TFIM}

\subsection{Implementation of expectation values}

For this derivation, mostly following \cite{dziarmaga_dynamics_2006}, we start with the bond-disordered TFIM for \(N=128\) spins with periodic boundary conditions. In this model, the spin-spin couplings \(J_i\) are generated from a uniform distribution \(J_i \in \left(0,2\right)\). For every calculation used here, 20 disorder realizations of all \(J_i\) are sampled, leading to 20 different Hamiltonian realizations. The quantities per disorder realization, which we will derive in the following, are averaged over all those disorder realization, and the standard error of the mean serves as a quantifier of the uncertainty. As the \(J_i\) are disordered, there is no discrete translation symmetry in this model, and hence the Hamiltonian will not fragment into subspaces that conserve the modulus of the wavevector \(k\). Instead, we use the most general form of the Bogoliubov transformation, involving all possible operators
\begin{equation}
    \gamma_m = u_{mn}^\ast c_n +v_{mn}^\ast c_n^\dagger,
    \label{eq:Bog_trans_gamma}
\end{equation}
where the prefactors \(u_{nm}\) and \(v_{nm}\) are complex numbers, and we used Einstein summation convention. In order for the \(\gamma_m\) to fulfill the fermionic commutation relations, the prefactors are subject to the constraints
\begin{align}
    \begin{cases}
        u_{mn} v_{ln} + v_{mn} u_{ln} = 0, \\
        u_{nl} u_{nm}^\ast + v_{nl} v_{nm}^\ast = \delta_{lm}.
    \end{cases}
    \label{eq:Bog_trans_gm}
\end{align}
Using these constraints, the reverse transformation reads
\begin{equation}
    c_n = u_{mn}\gamma_m + v_{mn}^\ast \gamma_m^\dagger.
    \label{eq:Bog_trans_dis}
\end{equation}
We insert this transformation after a Jordan-Wigner transformation of the bond-disordered TFIM and demand that the Hamiltonian can be represented by non-interacting Bogoliubov quasiparticles
\begin{equation}
    H = \sum_m \epsilon_m \left(\gamma_m^\dagger \gamma_m - \frac{1}{2 } \right),
    \label{eq:Ham_diag_qp}
\end{equation}
where \(\epsilon_m\) is the energy related to the creation of quasiparticle \(m\).
We rewrite the bond-disordered TFIM after a Jordan-Wigner transformation into the form
\begin{equation}
    H = \frac{1}{2}\begin{pmatrix}
        c_1^\dagger & \dots c_N^\dagger & c_1 & \dots & c_N
    \end{pmatrix} \begin{pmatrix}
        a & b\\ -b^\ast & -a^\ast
    \end{pmatrix} \begin{pmatrix}
        c_1 \\ \vdots \\ c_N \\ c_1^\dagger \\ \vdots \\ c_N^\dagger
    \end{pmatrix}    ,
    \label{eq:ham_diag_cn}
\end{equation}
where \(a\) is a Hermitian and \(b\) an antisymmetric \(N \times N\) matrix. In order for the transformation Eq. \ref{eq:Bog_trans_dis} to yield the Hamiltonian Eq. \ref{eq:Ham_diag_qp}, we arrive at the Bogoliubov-de Gennes equation
\begin{equation}
    \epsilon_m \begin{pmatrix}
        u_{mn} \\
        v_{mn}
    \end{pmatrix} = \begin{pmatrix}
        a & b \\ -b^\ast & -a^\ast
    \end{pmatrix} \begin{pmatrix}
        u_{mn} \\
        v_{mn}
    \end{pmatrix}
    \label{eq:Bog_dg_dis}
\end{equation}
We observe that the \(\begin{pmatrix}
    u_{mn} & v_{mn}
\end{pmatrix}\) are eigenvectors of a Hermitian matrix, and as such immediately fulfill Eq. \ref{eq:Bog_trans_gm}. In addition, we observe another symmetry: If \(\begin{pmatrix}
    u_{mn} & v_{mn}
\end{pmatrix}\) is an eigenvector with positive-energy \(\epsilon_m\), \(\begin{pmatrix}
    v_{mn}^\ast & u_{mn}^\ast
\end{pmatrix}\) is a solution with the negated energy \(-\epsilon_m\). From Eq. \ref{eq:Bog_trans_gamma} it can be seen that this corresponds to the annihilation of quasiparticle \(m\), and hence is associated with the quasiparticle energy of a hole. Thus, only the positive-energy quasiparticles represent the creation of quasiparticles, whereas the negative-energy quasiparticles represent their annihilation.

As in the case for the uniform TFIM, we can obtain the time-dependent Bogoliubov-de Gennes equation by applying the Heisenberg equation of motion 
\begin{equation}
    i \frac{dc_n}{dt} = \comm{c_n}{H}
    \label{eq:Heisenberg_cn}
\end{equation}
and applying the transformation Eq. \ref{eq:Bog_trans_dis}. Assuming the initial state in the Heisenberg picture was the vacuum state annihilated by all \(\gamma_m\), and using the \(\gamma_m\) as a time-independent generator of the fermion algebra, i.e. \(\frac{d\gamma_m}{dt}=0\), the time-dependent Bogoliubov-de Gennes equations read
\begin{equation}
    i \frac{d}{dt} \begin{pmatrix}
        u_{mn}(t) \\
        v_{mn}(t)
    \end{pmatrix} = \begin{pmatrix}
        a(t) & b(t) \\ -b^\ast (t) & -a^\ast (t)
    \end{pmatrix} \begin{pmatrix}
        u_{mn} (t) \\
        v_{mn} (t)
    \end{pmatrix}
    \label{eq:Bog_dg_dis_time}
\end{equation}
We write the matrices \(a\) and \(b\) here in general time dependent. In our case, b depends only on the spin-spin couplings and is thus time-independent, while \(a\) has a dependence on the magnetic field \(h\) and as such is time dependent.
For the reverse and forward quench protocols, the magnetic field follows the time dependence

\begin{align}
    \label{eq:time_dep_a}
    h(t) = \frac{\left|t\right|}{\tau_Q}, \quad -10\tau_Q \le t \le 10\tau_Q, 
    \\
    \label{eq:time_dep_b}
    h(t) = \frac{t}{\tau_Q},\quad  0 \le t \le 10 \tau_Q,
\end{align}

respectively. In order to calculate the time evolution for a system initially prepared in the ground state, we use Eq. \ref{eq:Bog_trans_gamma} to get the initial values \(u_{mn}(h_\mathrm{init}) \) and \(v_{mn}(h_\mathrm{init})\), where \(h_\mathrm{init}\) denotes the magnetic field at initial times.
We insert these initial conditions into Eq. \ref{eq:Bog_dg_dis_time}, using the time-dependence of the field in Eqs. \ref{eq:time_dep_a} and \ref{eq:time_dep_b} to obtain the values \(u_{mn}(t_\mathrm{final}) \coloneqq \tilde{u}_{mn}\) and \(v_{mn}(t_\mathrm{final})\coloneqq \tilde{v}_{mn}\), where \(t_\mathrm{final}\) denotes the time after the forward and reverse quench, respectively. The numerical solution was obtained using the QuantumOptics.jl package \cite{kramer_quantumopticsjl_2018}.
These values can be used to calculate the expectation value of the defect density at the final field \(h_\mathrm{final}\) at the end of the quench. Using Eqs. \ref{eq:Bog_trans_gamma} and \ref{eq:Bog_dg_dis}, we can rewrite the quasiparticle operator at the field \(h_\mathrm{final}\) as a function of time as 
\begin{equation}
    \gamma_m(t) = \bar{u}_{mn}^\ast c_n(t) +\bar{v}_{mn}^\ast c_n^\dagger (t)
\end{equation}
where the \(\bar{u}_{mn}\) and \(\bar{v}_{mn}\) are obtained from solving Eq. \ref{eq:Bog_dg_dis} restricting the solutions to positive-energy eigenstates with \(h=h_\mathrm{final}\).
Then, the quasiparticle defect density reads:
\begin{equation}
\begin{split}
    n_\mathrm{def} &= \frac{1}{N} \sum_{m=1}^N \gamma_m^\dagger(h_\mathrm{final},t) \gamma_m(h_\mathrm{final},t)\\
    &= \frac{1}{N} \sum_{m=1}^N \sum_{n=1}^N \sum_{p=1}^N (\bar{u}_{mn} c_n^\dagger(t)+\bar{v}_{mn}c_n(t)) \\& (\bar{u}_{mp}^\ast c_p(t)+\bar{v}_{mp}^\ast c_p^\dagger(t))
\end{split}  
\label{eq:exact_def_cn}
\end{equation}
Inserting the time-dependent result for the \(c_n(t)\) using Eq. \ref{eq:Bog_trans_dis}, we obtain for the defect density at final field \(h_\mathrm{final}\) at time \(t_\mathrm{final}\)
\begin{equation}
    n_\mathrm{def} = \frac{1}{N} \sum_m \begin{pmatrix}
        \tilde{u}_{ml}^\ast & \tilde{v}_{ml}^\ast
    \end{pmatrix} \begin{pmatrix}
        \bar{v}_{ql}^\ast \bar{v}_{qn} & \bar{v}_{ql}^\ast \bar{u}_{qn}\\
        \bar{u}_{ql}^\ast \bar{v}_{qn} & \bar{u}_{ql}^\ast \bar{u}_{qn}
    \end{pmatrix} \begin{pmatrix}
        \tilde{u}_{mn}\\ \tilde{v}_{mn}
    \end{pmatrix}
    \label{eq:def_disordered}
\end{equation}

The magnetization at the final time can be calculated in a similar way
\begin{equation}
\begin{split}
   & \expval{\hat{S}_x(t)} = \frac{1}{2N}\sum_{n=1}^N\expval{\sigma_n^x(t)}\\
    &= \frac{1}{2N}\sum_{n=1}^N\expval{1-2c_n^\dagger(t)c_n(t)}\\
   & = \frac{1}{2}\\
   &-\frac{1}{N} \sum_{m,n,l=1}^N \expval{(u_{mn}^\ast(t) \gamma_m^\dagger +v_{mn}(t)\gamma_m) (u_{ln}(t) \gamma_l +v_{ln}^\ast(t)\gamma_l^\dagger)}\\
  &  = \frac{1}{2} - \frac{1}{N}\sum_{m,n} \left|v_{mn}(t)\right|^2\\
  &= \frac{1}{2N} \sum_{m,n} \left(\left|u_{mn}(t)\right|^2+\left|v_{mn}(t)\right|^2\right) - \frac{1}{N}\sum_{m,n} \left|v_{mn}(t)\right|^2\\
  &= \frac{1}{2N} \sum_{m,n} \left(\left|u_{mn}(t)\right|^2-\left|v_{mn}(t)\right|^2\right)
\end{split}
\label{eq:mag_disordered}
\end{equation}
\subsection{Implementation of variances}
For simplicity of the calculation, we use the Einstein summation convention, and the same definition of \(\bar{u}_{mn}, \bar{v}_{mn}\) and \(\tilde{u}_{mn},\tilde{v}_{mn}\) as in the previous section. For a chain of  \(N\) spins, we obtain for the expectation value of the square of the total number of defects \(N_\mathrm{def}\):
\begin{equation}
\begin{split}
    \left \langle N_\mathrm{def}^2 \right \rangle &= \bra{0} \gamma_m^\dagger(h_\mathrm{final},t) \gamma_m(h_\mathrm{final},t) \gamma_n^\dagger(h_\mathrm{final},t) \gamma_n(h_\mathrm{final},t) \ket{0}\\
    &= \bra{0}  (\bar{u}_{mq} c_q^\dagger(t)+\bar{v}_{mq}c_q(t)) (\bar{u}_{mk}^\ast c_k(t)+\bar{v}_{mk}^\ast c_k^\dagger(t)) \\ & (\bar{u}_{nl} c_l^\dagger(t)+\bar{v}_{nl}c_l(t)) (\bar{u}_{np}^\ast c_p(t)+\bar{v}_{np}^\ast c_p^\dagger(t)) \ket{0}\\
    &= \bra{0} \left[ \bar{u}_{mq} \left(\tilde{u}_{aq}^\ast \gamma_a^\dagger +\tilde{v}_{aq} \gamma_a \right)+\bar{v}_{mq}\left(\tilde{u}_{aq} \gamma_a + \tilde{v}_{aq}^\ast \gamma_a^\dagger \right)\right]\\
    & \left[ \bar{u}_{mk}^\ast \left(\tilde{u}_{bk} \gamma_b +\tilde{v}_{bk}^\ast \gamma_b^\dagger \right)+\bar{v}_{mk}^\ast \left(\tilde{u}_{bk}^\ast \gamma_b^\dagger + \tilde{v}_{bk} \gamma_b \right)\right]\\
    & \left[ \bar{u}_{nl} \left(\tilde{u}_{cl}^\ast \gamma_c^\dagger +\tilde{v}_{cl} \gamma_c \right)+\bar{v}_{nl}\left(\tilde{u}_{cl} \gamma_c + \tilde{v}_{cl}^\ast \gamma_c^\dagger \right)\right]\\
    & \left[ \bar{u}_{np}^\ast \left(\tilde{u}_{dp} \gamma_d +\tilde{v}_{dp}^\ast \gamma_d^\dagger \right)+\bar{v}_{np}^\ast \left(\tilde{u}_{dp}^\ast \gamma_d^\dagger + \tilde{v}_{dp} \gamma_d \right)\right] \ket{0}\\
    &= \left(\bar{u}_{mq} \tilde{v}_{aq} + \bar{v}_{mq}\tilde{u}_{aq} \right) \left(\bar{u}_{mk}^\ast\tilde{v}_{bk}^\ast + \bar{v}_{mk}^\ast \tilde{u}_{bk}^\ast \right)\\
    &\left(\bar{u}_{nl}\tilde{v}_{cl}+\bar{v}_{nl}\tilde{u}_{cl}\right) \left(\bar{u}_{np}^\ast\tilde{v}_{dp}^\ast+\bar{v}_{np}^\ast\tilde{u}_{dp}^\ast\right) \underbrace{\expval{\gamma_a \gamma_b^\dagger \gamma_c \gamma_d^\dagger}{0}}_{=\delta_{ab}\delta_{cd}}\\
    & + \left(\bar{u}_{mq} \tilde{v}_{aq} + \bar{v}_{mq}\tilde{u}_{aq} \right) \left(\bar{u}_{mk}^\ast\tilde{u}_{bk} + \bar{v}_{mk}^\ast \tilde{v}_{bk}\right)\\ &\left(\bar{u}_{nl}\tilde{u}_{cl}^\ast+\bar{v}_{nl}\tilde{v}_{cl}^\ast\right) \left(\bar{u}_{np}^\ast\tilde{v}_{dp}^\ast+\bar{v}_{np}^\ast\tilde{u}_{dp}^\ast\right) \underbrace{\expval{\gamma_a \gamma_b \gamma_c^\dagger \gamma_d^\dagger}{0}}_{=\delta_{ad}\delta_{bc}-\delta_{ac}\delta_{bd}}\\
    &= \left(\bar{u}_{mq} \tilde{v}_{aq} + \bar{v}_{mq}\tilde{u}_{aq} \right) \left(\bar{u}_{mk}^\ast\tilde{v}_{ak}^\ast + \bar{v}_{mk}^\ast \tilde{u}_{ak}^\ast \right)\\
    &\left(\bar{u}_{nl}\tilde{v}_{cl}+\bar{v}_{nl}\tilde{u}_{cl}\right) \left(\bar{u}_{np}^\ast\tilde{v}_{cp}^\ast+\bar{v}_{np}^\ast\tilde{u}_{cp}^\ast\right)\\
    & + \left(\bar{u}_{mq} \tilde{v}_{aq} + \bar{v}_{mq}\tilde{u}_{aq} \right) \left(\bar{u}_{np}^\ast\tilde{v}_{ap}^\ast+\bar{v}_{np}^\ast\tilde{u}_{ap}^\ast\right)\\ &\left(\bar{u}_{mk}^\ast\tilde{u}_{bk} + \bar{v}_{mk}^\ast \tilde{v}_{bk}\right) \left(\bar{u}_{nl}\tilde{u}_{bl}^\ast+\bar{v}_{nl}\tilde{v}_{bl}^\ast\right)\\
    &- \left(\bar{u}_{mq} \tilde{v}_{aq} + \bar{v}_{mq}\tilde{u}_{aq} \right) \left(\bar{u}_{nl}\tilde{u}_{al}^\ast+\bar{v}_{nl}\tilde{v}_{al}^\ast\right)\\
    &\left(\bar{u}_{mk}^\ast\tilde{u}_{bk} + \bar{v}_{mk}^\ast \tilde{v}_{bk}\right) \left(\bar{u}_{np}^\ast\tilde{v}_{bp}^\ast+\bar{v}_{np}^\ast\tilde{u}_{bp}^\ast\right)
\end{split}  
\label{eq:exact_def_var_cn}
\end{equation}
A close comparison of the first summand in the last expression to Eq. \ref{eq:def_disordered} shows that this term is equal to \(\left \langle N_\mathrm{def} \right \rangle^2= \left \langle N n_\mathrm{def} \right \rangle^2\). As a consequence, the variance of the number of defects is given by
\begin{equation}
\begin{split}
    \mathrm{Var}\left(N_\mathrm{def}\right) & =\left \langle N_\mathrm{def}^2 \right \rangle -\left \langle N_\mathrm{def} \right \rangle^2\\
    &= \left(\bar{u}_{mq} \tilde{v}_{aq} + \bar{v}_{mq}\tilde{u}_{aq} \right) \left(\bar{u}_{np}^\ast\tilde{v}_{ap}^\ast+\bar{v}_{np}^\ast\tilde{u}_{ap}^\ast\right),\\ &\left(\bar{u}_{mk}^\ast\tilde{u}_{bk} + \bar{v}_{mk}^\ast \tilde{v}_{bk}\right) \left(\bar{u}_{nl}\tilde{u}_{bl}^\ast+\bar{v}_{nl}\tilde{v}_{bl}^\ast\right)\\
    &- \left(\bar{u}_{mq} \tilde{v}_{aq} + \bar{v}_{mq}\tilde{u}_{aq} \right) \left(\bar{u}_{nl}\tilde{u}_{al}^\ast+\bar{v}_{nl}\tilde{v}_{al}^\ast\right)\\
    &\left(\bar{u}_{mk}^\ast\tilde{u}_{bk} + \bar{v}_{mk}^\ast \tilde{v}_{bk}\right) \left(\bar{u}_{np}^\ast\tilde{v}_{bp}^\ast+\bar{v}_{np}^\ast\tilde{u}_{bp}^\ast\right).
\end{split}
\label{eq:variation_derivation}
\end{equation}
Calculating the defects involves knowledge of all eigenvectors \(\left(u,v\right)\) at the final field \(\left(\bar{u},\bar{v}\right)\) as well as the time evolved values \(\left(\tilde{u},\tilde{v}\right)\). In order to further simplify the expression, we define the matrices \(\mathbf{\bar{u}} \coloneqq \bar{u}_{mn}\), \(\mathbf{\bar{v}} \coloneqq \bar{v}_{mn}\), \(\mathbf{\tilde{u}} \coloneqq \tilde{u}_{mn}\) and \(\mathbf{\tilde{v}} \coloneqq \tilde{v}_{mn}\).

Then, Eq. \ref{eq:variation_derivation} can be rewritten in the form
\begin{equation}
\begin{split}
    \mathrm{Var}\left(N_\mathrm{def}\right) &= \left(\mathbf{\bar{u}}\mathbf{\tilde{v}}^\top+\mathbf{\bar{v}}\mathbf{\tilde{u}}^\top\right)_{ma} \left(\mathbf{\bar{u}}^\ast\mathbf{\tilde{v}}^\dagger+\mathbf{\bar{v}}^\ast\mathbf{\tilde{u}}^\dagger\right)_{na}\\
    & \left(\mathbf{\bar{u}}^\ast\mathbf{\tilde{u}}^\top+\mathbf{\bar{v}}^\ast\mathbf{\tilde{v}}^\top\right)_{mb} \left(\mathbf{\bar{u}}\mathbf{\tilde{u}}^\dagger+\mathbf{\bar{v}}\mathbf{\tilde{v}}^\dagger\right)_{nb}\\
    & - \left(\mathbf{\bar{u}}\mathbf{\tilde{v}}^\top+\mathbf{\bar{v}}\mathbf{\tilde{u}}^\top\right)_{ma} \left(\mathbf{\bar{u}}\mathbf{\tilde{u}}^\dagger+\mathbf{\bar{v}}\mathbf{\tilde{v}}^\dagger\right)_{na}\\
    & \left(\mathbf{\bar{u}}^\ast\mathbf{\tilde{u}}^\top+\mathbf{\bar{v}}^\ast\mathbf{\tilde{v}}^\top\right)_{mb} \left(\mathbf{\bar{u}}^\ast\mathbf{\tilde{v}}^\dagger+\mathbf{\bar{v}}^\ast\mathbf{\tilde{u}}^\dagger\right)_{nb}
\end{split}
\label{eq:variation_simple_matrices}
\end{equation}

\begin{figure*}[htb]
    \centering
    \includegraphics[width=0.9\linewidth]{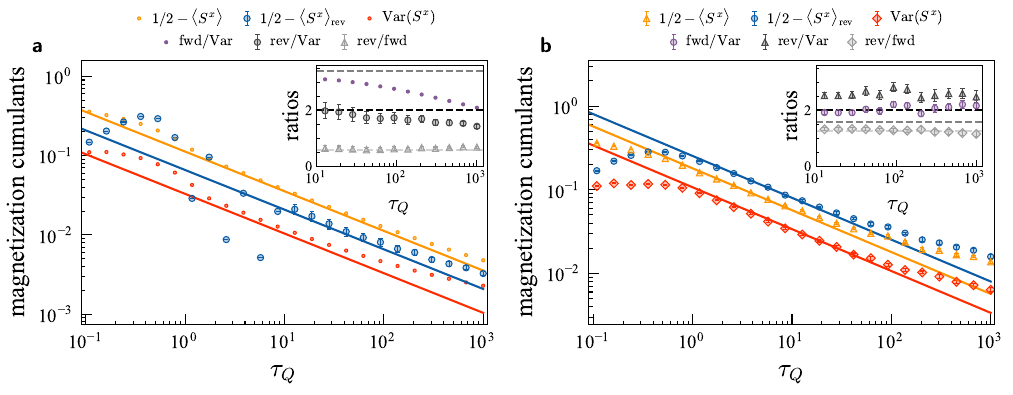}
    \caption{One half minus magnetization after a forward (orange) and reverse (blue) quench, and magnetization variance after a forward quench (red), versus quench time \(\tau_Q\) for the uniform (a) and bond-disordered (b) TFIMs. Solid lines in (a) show the power-law predictions for the exact defect densities. In (b), the lines illustrate the same power-law dependence \(1/2-\expval{\hat{S}_x}=C/(2\pi\sqrt{2\tau_Q})\), with \(C=1.6\), \(0.95\), and \(2.25\) for one half minus the magnetization after a forward quench, for the magnetization variance after a forward quench, and for one half minus the magnetization after a reverse quench, respectively. Error bars show the standard error of the mean, calculated from 11 evenly spaced points spanning two Stückelberg oscillations in the uniform model and from 20 disorder realizations in the bond-disordered model. Insets show, in violet, gray, and light gray, respectively, the ratios of the forward-quench (fwd) and reverse-quench (rev) values of one half minus the magnetization to the forward-quench magnetization variance (Var), and the ratio of the forward- to reverse-quench values. The black, gray, and light-gray dashed lines mark \(2\), the value \(1/(1-p)\), and \(2-2p\), respectively, where \(p = 1/\sqrt{2}\) for the uniform TFIM and \(p = 0.37\) for the bond-disordered TFIM.}
    \label{fig:mags}
\end{figure*}

In order to reduce the number of redundant calculations in a numerical calculation, we further introduce the matrices \(\mathbf{A}\coloneqq \mathbf{\tilde{v}}\mathbf{\bar{u}}^\top+\mathbf{\tilde{u}}\mathbf{\bar{v}}^\top\) and \(\mathbf{B}\coloneqq \mathbf{\bar{u}}^\ast\mathbf{\tilde{u}}^\top+\mathbf{\bar{v}}^\ast\mathbf{\tilde{v}}^\top\) and their product \(\mathbf{C} \coloneqq \mathbf{A}\mathbf{B}\). Then, equation \ref{eq:variation_simple_matrices} takes the simple form
\begin{equation}
    \mathrm{Var}\left(N_\mathrm{def}\right) = \Tr{\mathbf{C} \mathbf{C}^\dagger}-\Tr{\mathbf{C} \mathbf{C}^\ast}
    \label{eq:variance_exact}
\end{equation}
as a consequence, the density of the defect number variance is given as
\begin{equation}
    \mathrm{Var}\left(n_\mathrm{def}\right) \coloneqq \frac{\mathrm{Var}\left(N_\mathrm{def}\right)}{N} = \frac{\Tr{\mathbf{C} \mathbf{C}^\dagger}-\Tr{\mathbf{C} \mathbf{C}^\ast}}{N}
    \label{eq:variance_density_exact}
\end{equation}

Eq. \ref{eq:def_disordered} can also be converted into a trace, namely
\begin{equation}
n_\mathrm{def} = \frac{\Tr{\mathbf{A}^\dagger\mathbf{A}}}{N}
    \label{eq:def_disordered_matrix}
\end{equation}
In the limit of \(h_\mathrm{final} \rightarrow \infty\), \( \hat{S}_x = 1/2-\hat{n}_\mathrm{def}\) applies. In this limit, we obtain \(\mathbf{\bar{u}}=\mathbb{1}_{N \times N}\) and \(\mathbf{\bar{v}}=0\). As a consequence, the expressions for \(\mathbf{A}\) and \(\mathbf{B}\) simplify to \(\mathbf{A}= \mathbf{\tilde{v}}\) and \(\mathbf{B} = \mathbf{\tilde{u}}^\top\). Inserting these into Eqs. \ref{eq:variance_density_exact} and \ref{eq:def_disordered_matrix} we obtain for the magnetization and the average magnetization fluctuation:
\begin{align}
    \left \langle \hat{S}_x \right \rangle = \frac{1}{2}-\frac{1}{N}\Tr{\mathbf{\tilde{v}}^\dagger \mathbf{\tilde{v}}},\\
    \mathrm{Var}\left(\hat{S}_x \right) = \frac{\Tr{\mathbf{\tilde{v}}\mathbf{\tilde{u}}^\top \mathbf{\tilde{u}}^\ast \mathbf{\tilde{v}}^\dagger}-\Tr{\mathbf{\tilde{v}}\mathbf{\tilde{u}}^\top \mathbf{\tilde{v}}^\ast\mathbf{\tilde{u}}^\dagger}}{N}.
    \label{eq:disordered_mag_var}
\end{align}

\section{Discussion of magnetization dynamics}

For the same values of \(h_0\) and \(\tau_Q\) as in Fig. 2 in the main text, we report the numerically simulated values of \(1/2 - \left\langle \hat{S}_x \right \rangle\) following forward and reverse quenches, as well as the variance of the magnetization following a forward quench, as functions of the quench time for the uniform and bond-disordered TFIM.

In the uniform model, the forward and reverse quench values of
\(\frac{1}{2} - \left\langle \hat{S}_x \right\rangle\) follow the analytic predictions for the defect densities well, even for \(h_0 = 10\) and a finite system of \(10^4\) particles. Within a \(1\sigma\) confidence interval, \(\frac{1}{2} - \left\langle \hat{S}_x \right\rangle_{\mathrm{rev}}\) approximates \(n_{\mathrm{def}}^{\mathrm{rev}}\) for \(20 \leq \tau_Q \leq 140\); over the same interval, \(\frac{1}{2} - \left\langle \hat{S}_x \right\rangle\) approximates \(n_{\mathrm{def}}\) to within a few percent. Moreover, the ratio of the forward  and reverse quench values of \(\frac{1}{2} - \left\langle \hat{S}_x \right\rangle\) is constant over this interval and is consistent with the analytic prediction.

By contrast, \(\mathrm{Var} \left(\hat{S}_x\right)\) does not provide a good approximation to \(\mathrm{Var}(n_{\mathrm{def}})\) and is more sensitive to finite magnetic fields. Consequently, the ratio of \(\frac{1}{2} - \left\langle \hat{S}_x \right\rangle_{\mathrm{rev}}\) to \(\mathrm{Var} \left(\hat{S}_x\right)\) quickly falls below the theoretically predicted value of \(2\), while the ratio of \(\frac{1}{2} - \left\langle \hat{S}_x \right\rangle\) to \(\mathrm{Var} \left(\hat{S}_x\right)\) varies with \(\tau_Q\).

\begin{figure*}
    \centering
    \includegraphics[width=0.9\linewidth]{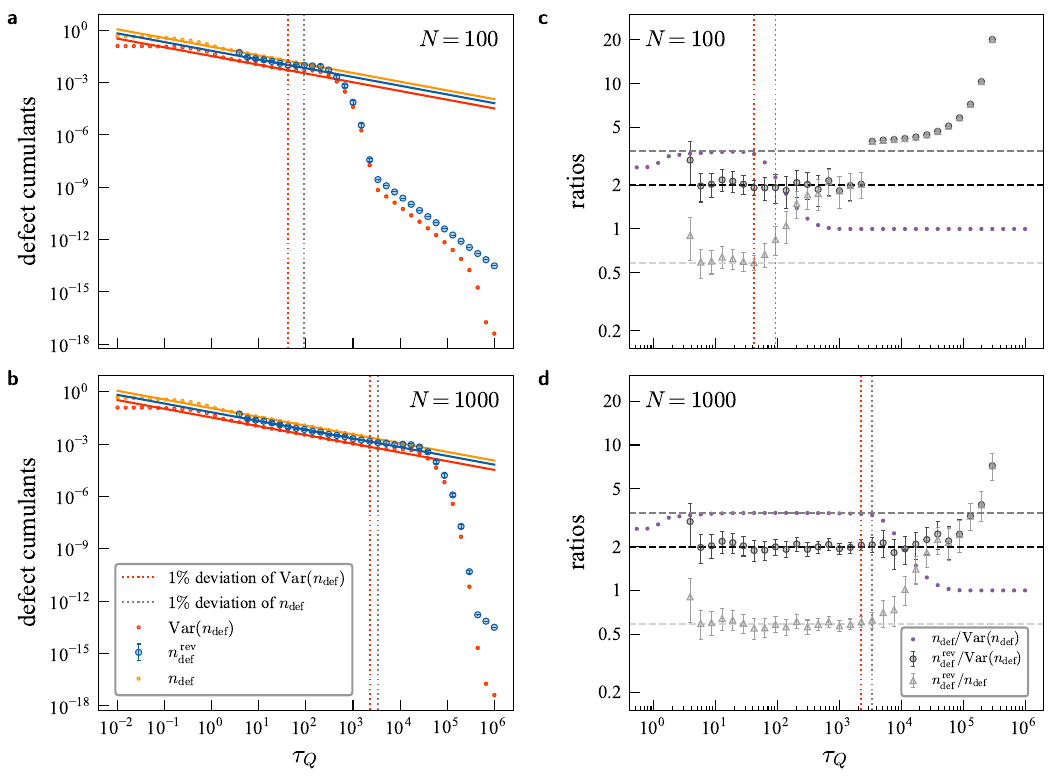}
    \caption{Defect statistics of the transverse-field Ising chain for system sizes \(N = 100\) (a, d), \(N = 1000\) (b, e), and \(N = 10000\) (c, f).
Left column (a--c): defect density after the reverse quench, \(n^\mathrm{rev}_\mathrm{def}\) (blue), and mean \(n_\mathrm{def}\) (yellow) and variance \(\mathrm{Var}(n_\mathrm{def})\) (red) of the defect density after the forward quench, as functions of the quench time \(\tau_Q\). Solid lines are the analytic predictions in the thermodynamic limit, \(n_\mathrm{def} = 1/(2\pi\sqrt{2\tau_Q})\), \(n^\mathrm{rev}_\mathrm{def} = (2-\sqrt{2})\, n_\mathrm{def}\), and \(\mathrm{Var}(n_\mathrm{def}) = (1-1/\sqrt{2})\, n_\mathrm{def}\).
Right column (d--f): the ratios \(n_\mathrm{def}/\mathrm{Var}(n_\mathrm{def})\) (violet), \(n^\mathrm{rev}_\mathrm{def}/\mathrm{Var}(n_\mathrm{def})\) (dark gray), and \(n^\mathrm{rev}_\mathrm{def}/n_\mathrm{def}\) (light gray).
Dashed horizontal lines mark the analytic values \(1/(1-p)=1/(1-1/\sqrt{2}) \approx 3.41\) (dark gray), \(2\) (black), and \(2(1-p) = 2-\sqrt{2} \approx 0.59\) (light gray), respectively.
Dotted vertical lines mark the first simulated \(\tau_Q\) at which \(n_\mathrm{def}\) (gray) and \(\mathrm{Var}(n_\mathrm{def})\) (red) deviate by more than \(1\%\) from the analytic prediction. For \(N = 10000\) the two lines coincide.
Each reverse-quench data point is averaged over 11 evenly spaced values of \(\tau_Q\) spread evenly within \(\pm\pi\) of the nominal value, and error bars give the standard error of the mean.
}
    \label{fig:finitesize}
\end{figure*}

All predicted values follow the \(\tau_Q^{-1/2}\) scaling for intermediate times, while for late times, the values show a slower scaling due to the finite magnetic field.

However, comparing \(\frac{1}{2} - \left\langle \hat{S}_x \right\rangle_{\mathrm{rev}}\) to \(\frac{1}{2} - \left\langle \hat{S}_x \right\rangle\) shows that their ratio takes the analytically predicted value of \(2-\sqrt{2}\) for all simulated values of \(\tau_Q\) within a \(1\sigma\) confidence interval for \(10 \le \tau_Q\le 300\), and within a \(2\sigma\) confidence interval for all simulated times.

In the bond-disordered model, we find essentially the same behavior. While the values of \(\frac{1}{2} - \left\langle \hat{S}_x \right\rangle\) and \(1/2 - \left\langle \hat{S}_x \right\rangle_\mathrm{rev}\) approximate the values of \(n_\mathrm{def}\) and \(n_\mathrm{def}^\mathrm{rev}\), respectively, well, the magnetization variance \(\mathrm{Var} \left(\hat{S}_x\right)\) drops significantly faster than \(\mathrm{Var}(n_\mathrm{def})\), thus being more sensible to finite magnetic field effects. As a consequence, the ratio between one half minus the magnetization both after a forward and a reverse quench to the magnetization quench variance are significantly larger than the numerically simulated values of the ratios of the defect density cumulants.
On the other hand, the ratio between one half minus the magnetization after a reverse and a forward quench agree within a \(1\sigma\) confidence interval for \(60\le\tau_q\le670\), and for all simulated values within a \(2\sigma\) confidence interval. Hence, we find consistently within both the uniform and the bond-disordered TFIM that the magnetization variance is more sensitive to finite size and finite magnetic field effects than the expectation values. 

This makes magnetization measurements a more reliable means of extracting the defect density than measurements of the magnetization variance to extract the defect density variance.

\section{Finite size effects in the uniform TFIM model}
\label{sec:finite_size}

To assess finite-size effects, we repeat the calculations for the uniform TFIM presented in the main text for two smaller system sizes, \(N=100\) and \(N=1000\). For these sizes, the evolution remains tractable within the limits of the Helix cluster up to \(\tau_Q=1\times 10^6\), allowing us to access the finite-size-induced adiabatic regime.

We also test the assumption that fast oscillations can be averaged out by averaging the reverse-quench defect density over 11 evenly spaced points spanning four oscillation periods, rather than two, for each \(\tau_Q\). The sampling points are in the interval \([\tau_Q-\pi,\tau_Q+\pi]\). For the calculation of the double ramp defect density, we thus only consider quench times with \(\tau_Q>\pi\).

Fig. \ref{fig:finitesize}(a) and (b) shows \(n_\mathrm{def}\), \(n_\mathrm{def}^\mathrm{rev}\), and \(\mathrm{Var}(n_\mathrm{def})\) for 47 logarithmically spaced values of \(\tau_Q\) between \(10^{-2}\) and \(10^6\). For \(\tau_Q \approx 8\), all three observables already agree well with their thermodynamic-limit analytic predictions. At longer quench times, the dynamics cross over to the adiabatic regime. We characterize this crossover using two threshold times: the first \(\tau_Q>10\) at which \(n_\mathrm{def}\) deviates from its analytic prediction by more than \(1\%\) (gray dotted vertical line), and the first \(\tau_Q>10\) at which \(\mathrm{Var}(n_\mathrm{def})\) deviates by more than \(1\%\) (red dotted vertical line). As expected, the former threshold occurs at a shorter quench time than the latter, since variances are more susceptible to finite-size effects than expectation values. Both thresholds shift to longer quench times as the system size increases, indicating that they mark the onset of significant finite-size effects.

Fig. \ref{fig:finitesize} (c) and (d) shows the ratios of the three cumulants discussed above for system sizes \(N=100\) and \(N=1000\), respectively. In both cases, the ratios depart from their analytically predicted values in the thermodynamic limit for large quench times in both the transient and finite-size-dominated regimes. Notably, as finite-size effects set in, the ratios \(n_\mathrm{def}/\mathrm{Var}(n_\mathrm{def})\) and \(n_\mathrm{def}^\mathrm{rev}/n_\mathrm{def}\) deviate significantly from their analytic predictions, whereas \(n_\mathrm{def}^\mathrm{rev}/\mathrm{Var}(n_\mathrm{def})\) remains consistent with the predicted value of \(2\). This behavior suggests that, in this model, the crossover to the adiabatic regime is initially accompanied by a dependence of the defect-formation probability \(p\) on \(\tau_Q\), before the gKZM picture breaks down and the dynamics become adiabatic.

This behavior contrasts with that shown in Fig. 2 of the main text, where \(n_\mathrm{def}^\mathrm{rev}/n_\mathrm{def}\) exhibits a weaker finite-size dependence than the other ratios. This suggests that, in that model, the crossover to adiabatic dynamics is not accompanied by a time-dependent defect-formation probability \(p\). Nevertheless, in both models, the regime dominated by neither transient dynamics nor finite-size effects is precisely the regime in which all three ratios are constant and \(n_\mathrm{def}^\mathrm{rev}/\mathrm{Var}(n_\mathrm{def})=2\).

\section{Average vs. sinusoidal fit on reverse quench defect density estimation}

\begin{figure*}
    \centering
    \includegraphics[width=0.9\linewidth]{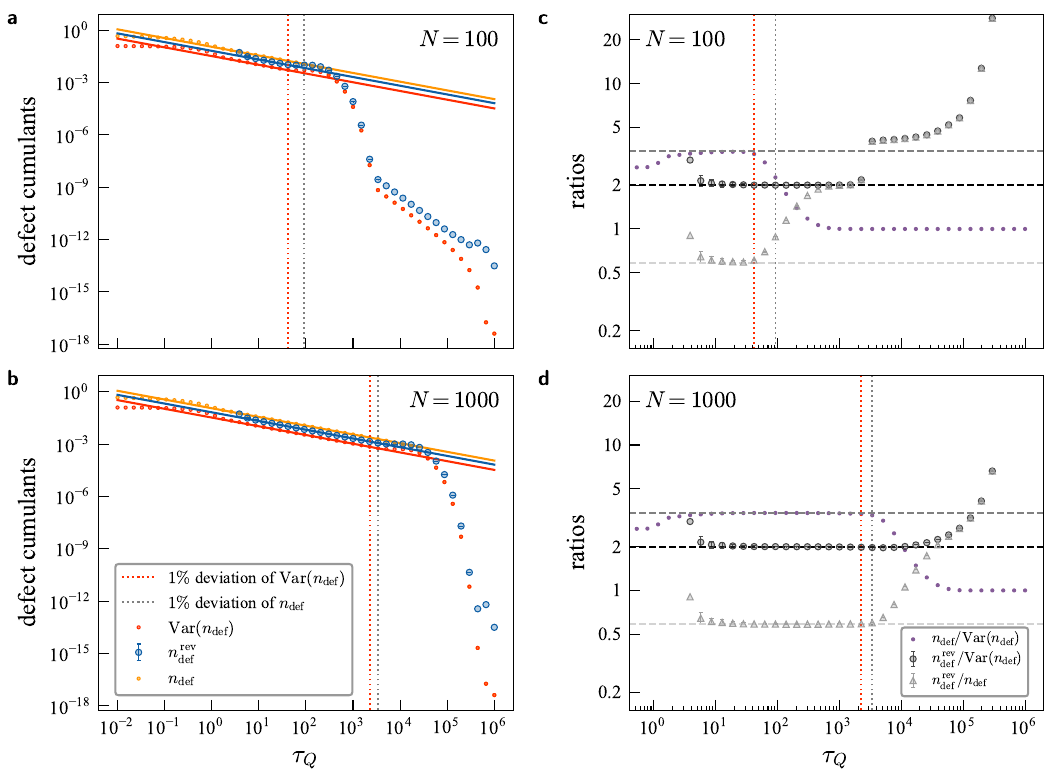}
    \caption{Defect statistics of the transverse-field Ising chain for system sizes \(N = 100\) (a, d), \(N = 1000\) (b, e), and \(N = 10000\) (c, f).
Left column (a--c): defect density after the reverse quench, \(n^\mathrm{rev}_\mathrm{def}\) (blue), and mean \(n_\mathrm{def}\) (yellow) and variance \(\mathrm{Var}(n_\mathrm{def})\) (red) of the defect density after the forward quench, as functions of the quench time \(\tau_Q\). Solid lines are the analytic predictions in the thermodynamic limit, \(n_\mathrm{def} = 1/(2\pi\sqrt{2\tau_Q})\), \(n^\mathrm{rev}_\mathrm{def} = (2-\sqrt{2})\, n_\mathrm{def}\), and \(\mathrm{Var}(n_\mathrm{def}) = (1-1/\sqrt{2})\, n_\mathrm{def}\).
Right column (d--f): the ratios \(n_\mathrm{def}/\mathrm{Var}(n_\mathrm{def})\) (violet), \(n^\mathrm{rev}_\mathrm{def}/\mathrm{Var}(n_\mathrm{def})\) (dark gray), and \(n^\mathrm{rev}_\mathrm{def}/n_\mathrm{def}\) (light gray).
Dashed horizontal lines mark the analytic values \(1/(1-p)=1/(1-1/\sqrt{2}) \approx 3.41\) (dark gray), \(2\) (black), and \(2(1-p) = 2-\sqrt{2} \approx 0.59\) (light gray), respectively.
Dotted vertical lines mark the first simulated \(\tau_Q\) at which \(n_\mathrm{def}\) (gray) and \(\mathrm{Var}(n_\mathrm{def})\) (red) deviate by more than \(1\%\) from the analytic prediction. For \(N = 10000\) the two lines coincide.
Each reverse-quench data point is fitted a sinusoidal over 11 evenly spaced values of \(\tau_Q\) spread evenly within \(\pm\pi\) of the nominal value, and error bars give one standard deviation of the fitted value, estimated from the covariance matrix.
}
\label{fig:oscillations}
\end{figure*}

So far, to estimate the reverse quench defect density, we took an average over the eleven linearly spaced data points around a single value of \(\tau_Q\), and took the standard error of the mean to estimate the error bar. Here, we want to discuss the pros and cons of a different approach, namely to fit the data a sinusoidal function of angular frequency 4 with a time dependent offset
\begin{equation}
    n_\mathrm{def}^\mathrm{rev}(\tau_Q+t) = A \sin \left(\omega t+\phi\right)+ n_\mathrm{def}^\mathrm{rev}(\tau_Q).
    \label{eq:fitfunc}
\end{equation}
where \(t\in \left[-\pi, -4\pi/5,\dots,4\pi/5, \pi\right]\), and from the analysis of the Stückelberg oscillations, we give \(\omega\) the initial value of \(\bar{omega}=4\) for a the fit, where the bar denotes the initial guess value.
Assuming that the underlying data indeed represent a sinusoidal of frequency 4 with a constant offset, we can get a very good estimate of the phase \(\phi\) and the offset \(n_\mathrm{def}^\mathrm{rev}(\tau_Q)\) by using 
\begin{equation}
    \bar{n}_\mathrm{dev}^\mathrm{rev}(\tau_Q) = \sum_{j=-5}^5 \frac{n^\mathrm{rev}_\mathrm{dev}(\tau_Q+j\pi/5)}{11}.
\end{equation}
Thus, the reduced value
\begin{equation}
    \tilde{n}_\mathrm{def}^\mathrm{rev}(\tau_Q+t) = n_\mathrm{def}^\mathrm{rev}(\tau_Q+t)- \bar{n}_\mathrm{dev}^\mathrm{rev}(\tau_Q)
\end{equation}
According to Eq. \ref{eq:fitfunc}, this should have the form of a pure sinusoidal function, which we may rewrite as
\begin{equation}
\begin{split}
    \tilde{n}_\mathrm{def}^\mathrm{rev}(\tau_Q+t) &= A \sin(\omega t+\phi)\\
    &= A\sin(\omega t)\cos(\phi)+A\cos(\omega t)\sin(\phi).
\end{split}
\end{equation}
Under the assumption that \(\omega=4\), the values of \(t\) span exactly four oscillation periods, and thus we obtain
\begin{align*}
    \frac{11}{2}\sin(\bar{\phi}) &= A = \sum_{j=-5}^5 \tilde{n}_\mathrm{def}^\mathrm{rev}(\tau_Q+t)\cos(4t),\\
    \frac{11}{2}\cos(\bar{\phi}) &= B = \sum_{j=-5}^5 \tilde{n}_\mathrm{def}^\mathrm{rev}(\tau_Q+t)\sin(4t),\\
    \bar{\phi} &= \arctan\left(\frac{A}{B}\right),
\end{align*}
where the \(\arctan\) operator takes into account the signs of both A and B.
The amplitude is initialized at \(\bar{A}=1\) and is constrained to positive values. The phase is restricted to \(\phi\in[-\pi,\pi]\), while the angular frequency is restricted to \(\omega\in[0,\omega_{\mathrm{Nyq}}]=[0,5]\), with \(\omega_{\mathrm{Nyq}}=\pi/\Delta t\) being the Nyquist frequency for our sampling points. This restriction confines the fit to the Nyquist interval and avoids assigning the signal to a frequency above the Nyquist limit.

The fit performed is a least-squares fit using the LMFIT python package \cite{newville_lmfit_2025}.

Fig. \ref{fig:oscillations} shows the same data as Fig. \ref{fig:finitesize}, except for the reverse quench defect densities being now fitted values, and their uncertainty is estimated from the covariance matrix of the fit.

At the earliest time point, \(\tau_Q \approx 3.8\), the covariance matrix cannot be reliably estimated for either of the two system sizes considered, so no error bar is shown. This is because, at such early times, the data are not expected to follow a sinusoidal form with a fixed frequency and offset. At intermediate times, we find excellent agreement with the analytical predictions. For very large quench times, as the system approaches the adiabatic limit, we do not expect oscillations to persist; consequently, the covariance matrix again does not provide a reliable uncertainty estimate. 

In this regime, the defect density is expected to scale approximately as \(\tau_Q^{-2}\) due to the finite many-body gap at finite system size \cite{schmitt_quantum_2022}. For \(\tau_Q \gtrsim 2 \times 10^5\), this scaling varies only weakly over an interval of length \(2\pi\), and therefore appears nearly constant. This is particularly evident in the fit values obtained in this regime: the fit does not converge, and, unlike the average over the 11 points, the fitted reverse-quench defect density exhibits non-monotonicities that are absent from the averaged values shown in Fig. \ref{fig:finitesize}. This indicates that the fit is unreliable at late times, which is the reason we preferred averaging for the data in the main text.

In Fig.~\ref{fig:oscillations}(c) and (d), we show the three ratios also presented in Fig.~\ref{fig:finitesize}(c) and (d) for system sizes \(N=100\) and \(N=1000\), respectively. The only difference we observe is that, in the intermediate regime where the gKZM holds and \(n_\mathrm{def}^\mathrm{rev}/\mathrm{Var}(n_\mathrm{def})=2\), the uncertainty in the reverse quench defect density and thus in the ratios involving it is significantly smaller. Nonetheless, our main findings remain unchanged: the gKZM does not hold in either the fast quench transient regime or the large quench time adiabatic regime. It holds only in the region where the defect cumulants follow their thermodynamic-limit power law scaling. Thus, the numerical data hint that the region where all three ratios are constant and \(n_\mathrm{def}^\mathrm{rev}/\mathrm{Var}(n_\mathrm{def})=2\) may serve an experimental signature that thermodynamic limit criticality is being observed.
\FloatBarrier

\bibliographystyle{apsrev4-2}
\bibliography{bib}